\documentclass[twocolumn]{revtex4}

\usepackage{graphicx}
\usepackage{epsf}
\usepackage{amsmath,amssymb}
\usepackage{ulem}
\usepackage{color}
\newcommand{\bvec}{\boldsymbol}
\newcommand{\braket}[1]{\langle #1\rangle}

\begin{document}

\newcommand{\kbar}{{K}}

\newcommand{\newlr}{\textrm{LR}}
\newcommand{\newrl}{\textrm{RL}}
\newcommand{\newll}{\textrm{LL}}
\newcommand{\newrr}{\textrm{RR}}

\newcommand{\newrt}{\textrm{R}}
\newcommand{\newlf}{\textrm{L}}
\newcommand{\newlfrt}{\textrm{L,R}}

\title{Effective interactions between nuclear clusters}

\author{Yoshiko Kanada-En'yo}
\affiliation{Department of Physics, Kyoto University, Kyoto 606-8502, Japan}
\author{Dean Lee}
\affiliation{Facility for Rare Isotope Beams and Department of Physics and Astronomy, Michigan State University, East Lansing, MI 48824, USA}

\begin{abstract}
The effective interactions between two nuclear clusters, $d+d$, $t+t$, and $\alpha+\alpha$, are investigated 
within a cluster model using local nucleon-nucleon~($NN$) forces. 
It is shown that the interaction 
in the spin-aligned $d+d$ system is repulsive 
for all inter-cluster distances, 
whereas the $\alpha+\alpha$ and spin-aligned $t+t$ systems 
are attractive at intermediate distances. 
The Pauli blocking between identical-nucleon pairs
is responsible for the cluster-cluster repulsion 
and becomes dominant in the shallow binding limit.  We demonstrate that two $d$-clusters could
be bound if the $NN$ force has nonzero range and is strong enough to form a deeply bound $d$-cluster,
or if the $NN$ force has both even-parity and odd-parity attraction.
Effective dimer-dimer interactions for general quantum systems of two-component fermions
are also discussed in heavy-light mass limit, where one component is much heavier than the other, and 
their relation to inter-cluster interactions in nuclear systems are discussed. Our findings provide a conceptual foundation for conclusions obtained
numerically in the literature, that increasing
the range or strength of the local part of the attractive nucleon-nucleon interaction
results in a more attractive cluster-cluster interaction.

\end{abstract}

\maketitle

\section{Introduction}
Nuclear clustering is a fascinating and important feature of many nuclear systems. Developed cluster structures appear in excited states of 
several nuclei and also in the ground states of systems such as 
$2\alpha$ clustering in $^8$Be$(0^+_1)$ and $^{16}$O+$\alpha$ 
clustering in $^{20}$Ne~\cite{Fujiwara80,Horiuchi:2012}. While
$\alpha$ clusters are the most common type of   cluster structure,
deuteron and triton clusters have also been suggested in light $p$-shell nuclei and at the surface of closed shell core nuclei.   
In highly excited states,
cluster states containing more than two clusters 
such as  $3\alpha$ structures in $^{12}$C and
$4\alpha$ structures in $^{16}$O 
have been attracting great interest 
in theoretical and experimental studies \cite{Fujiwara80,Horiuchi:2012,Freer:2014qoa,Freer:2017gip}.

The formation of clusters has been also investigated at the nuclear surface of 
$sd$- and heavier nuclei where spatial cluster correlations beyond mean-field may emerge \cite{Astier:2009bs,Ren:2018xpt}.  Concerning a two-nucleon pair with a strong spatial correlation, deuteron-like $pn$ and dineutron $nn$ correlations are also recent hot topics. 
For the latter, two neutrons are not bound in a free space, but the $nn$ correlation
is rather strong in  loosely-bound neutron-rich systems 
such as $^6$He and $^{11}$Li and can be regarded as a $(nn)$-cluster~\cite{Bertsch:1991zz,Zhukov:1993aw,Barranco:2000ip,Myo:2002wq,Hagino:2005we}. 
The possibility of an $\alpha+nn+nn$ structure has been proposed for 
an excited state of $^8$He \cite{Kanada-Enyo:2007iri}. 
Another candidate for multi-dineutron systems is
$nn+nn$ clustering in a four-neutron system called the tetraneutron. But this remain a controversial issue:
experimental signals of a tetraneutron resonance have been  
recent reported~\cite{Marques:2001wh,Kisamori:2016jie}
while several theoretical studies are not able to accommodate such a resonance \cite{Giraud:1973zz,Bertulani:2002px,Pieper:2003dc,Lazauskas:2005ig,Lashko:2006rp,
Hiyama:2016nwn,Fossez:2016dch}.

The effective interactions between clusters play an important role 
in cluster phenomena in nuclear systems. For example, the ground state of $^8$Be is 
a quasi-bound $2\alpha$ state formed by a short-range repulsion and a medium-range attraction 
of the effective $\alpha$-$\alpha$ interaction, which has been experimentally
observed from the $\alpha$-$\alpha$ scattering phase shifts. 
This $\alpha$-$\alpha$ interaction also
describes  the $3\alpha$ structure of the Hoyle state, $^{12}$C($0^+_2$).
The short-range repulsion and medium-range attraction, which are experimentally
known from the scattering phase shifts, are essential to describe the developed $3\alpha$ structure in $^{12}$C($0^+_2$). 
In a microscopic $\alpha+\alpha$ cluster model with the resonating group method (RGM),  
the repulsive effect of the $\alpha$-$\alpha$ interaction 
was described by a nodal structure of the inter-cluster wave function caused by 
the Pauli repulsion between identical nucleons in different clusters \cite{tamagaki68}. A similar Pauli effect
contributes to the effective interaction between two dineutrons
and produces significant repulsion in the tetraneutron system~\cite{Giraud:1973zz,Bertulani:2002px}.

In Ref.~\cite{Elhatisari:2016owd} it was observed that the $\alpha$-$\alpha$ interaction determines whether nuclear matter forms a nuclear liquid or a Bose-Einstein condensate (BEC) of alpha particles.  First principles calculations showed that the range and strength of the local part of the nucleon-nucleon interaction were essential for overcoming the Pauli blocking repulsion between the $\alpha$ particles \cite{Elhatisari:2016owd,Elhatisari:2015iga}. Here the term ``local interaction'' refers to an interaction kernel that is diagonal in the particle positions.  These results show that cluster-cluster interactions
are important not only for understanding specific nuclear states
with well-defined cluster substructures, but also important for understanding the balance of attractive and repulsive forces in nuclear matter.

Nuclear clustering is characterized by spatial correlations of the nucleons, and there are clear analogies to universal phenomena in other quantum degenerate fermionic systems. Dineutron correlations can be understood in terms of the universal properties of two-component fermionic superfluids at large scattering length \cite{Braaten:2004rn,Matsuo:2005vf,Giorgini:2008zz}, and $\alpha$ condensation in nuclear matter can be related to the general theory of fermionic quartet condensation \cite{Ropke:1998qs,Lee:2007eu,Ebran:2019rux}.
To understand the fundamental features of nuclear clustering and cluster-cluster interactions, it is useful to start with the dimer-dimer system. 
The dimer is the simplest composite system, having only two constituent particles.

In the limit of large particle-particle scattering length, the short-distance details of the interactions become irrelevant. In this universal limit we can simplify the particle-particle interactions to take the form of an attractive zero-range or delta-function interaction, taking care to properly renormalize the strength in the zero-range limit.  For two-component fermions in the limit of large scattering length, the dimer-dimer interaction is repulsive with a scattering length equal to 0.60 times the particle-particle scattering length \cite{Petrov2005,Elhatisari:2016hui,Deltuva:2017qgh}.

Recently, a study of effective dimer-dimer interactions  
for two-component fermions with general fermion-fermion interactions was performed using one-dimensional 
lattice calculations \cite{Rokash:2016tqh}. This study found repulsive dimer-dimer interactions for short-range forces but 
attractive dimer-dimer interactions for forces with larger range. 
It also found that local fermion-fermion interactions produced more attraction for the dimer-dimer interaction than nonlocal fermion-fermion interactions.  

 The universal repulsion for the dimer-dimer interaction at large scattering length appears also  
in mass imbalanced systems, where the two fermion components have masses $M$ and $m$ with $M > m$.  We find that this approach is useful for understanding the competition between attractive and repulsive forces analytically in the limit $M\gg m$, and we will refer to it as the heavy-light ansatz or Born-Oppenheimer approximation \cite{Lee:2007ae}. 
Questions to be answered are whether the nuclear force behaves as a short-range force, thus
producing universal repulsion between two dimers, and, if so, how the attractive $\alpha$-$\alpha$ interaction 
forms as the number and binding of the constituent nucleons within the clusters increase. 

In this work, we start with a general discussion of the effective dimer-dimer interactions
using the heavy-light ansatz and consider the relation to the effective 
inter-cluster interaction for the spin-aligned
$d+d$ system, which can be viewed as a two-dimer system composed of two-component
fermions with components corresponding to isospin.
We then investigate the effective inter-cluster interactions of $d+d$, $t+t$, and $\alpha+\alpha$ systems with a microscopic cluster model using Brink-Bloch two-cluster wave functions~\cite{brink66}
with effective nucleon-nucleon ($NN$) forces. 
We find a repulsive interaction in the spin-aligned $d+d$ system, attractive interactions in  
the spin-aligned $t+t$ and $\alpha+\alpha$ systems, and strong attractive interactions 
in the spin-opposed $d+d$ and $t+t$ systems.
By analyzing single-particle orbitals in the two-cluster systems, the impact of 
antisymmetrization between identical nucleons on the cluster-cluster interaction is illuminated.
Energies of the lowest states of two-cluster systems
are calculated with the generator coordinate method (GCM)~\cite{GCM1,GCM2}. 

The paper is organized as follows. In the next section, 
two-dimer systems with the heavy-light ansatz are described and 
effective dimer-dimer interactions are discussed. In Sec.~\ref{sec:two-cluster}, 
effective interactions between two clusters in 
nuclear systems are investigated. A summary is given in 
Sec.~\ref{sec:summary}. 
Appendix~\ref{app:two-delta} gives 
solutions of the two-delta potential problem in one dimension, 
and Appendix~\ref{app:volkov} describes 
parametrization of the effective $NN$ force.
Inter-cluster wave functions in two-cluster systems are described in Appendix~\ref{app:relative-wf}.


\section{Effective interaction between two dimers}\label{sec:heavy-light}

\subsection{Heavy-light ansatz $M\gg m$}
We consider a mass imbalanced system of two-component fermions, where the two fermion components have masses
$M$ and $m$ with $M > m$.  We assume an attractive and local $Mm$ potential that produces a bound $Mm$ dimer
and no interaction between identical particles.
We consider the limit $M\gg m$, and we call the resulting simplifications the heavy-light ansatz. The discussion will begin with the one-dimensional case, but will then move to the three-dimensional case soon afterwards. 

The heavy particles are stationary at coordinates at 
$\{{R}_1, {R}_2, \ldots\},$ and the light particles are feel the potentials produced by the heavy particles.  The Hamiltonian is
\begin{align}
&H=\sum_i h(i),\quad h(i)\equiv t(i)+U(i),\\
&t(i)=-\frac{\hbar^2}{2m}\frac{\partial^2}{\partial x^2_i},\quad U(i)=\sum_j v(|x_i-R_j|),
\end{align}
where $U$ is the one-body potential.  The ground state is a Slater determinant of single particle states,
\begin{align}
\Psi(1,\ldots,A_m) &={\cal A}\left\{\psi_1 \cdots \psi_{A_m}\right\}\nonumber\\
&=\frac{1}{\sqrt{A_m!}}\det\left\{\psi_1 \cdots \psi_{A_m}\right\}.
\end{align} 
$A_m$ is the total number of light $m$-particles and 
${\cal A}$ is the antisymmetrizer, and the single-particle states are
\begin{equation}
h(i)\psi_n(i)=e_n \psi_n(i).
\end{equation}
We here use the notation for the 
one-dimensional (1D) system, but it can be readily applied to the three dimensional (3D) problem by replacing
$x\to \bvec{x}$ and $R \to \bvec{R}$. It is also straightforward 
to extend the model to a nonlocal $Mm$ interaction.

For the single-dimer system ($Mm$), 
the Hamiltonian and wave function 
are given as 
\begin{align}
&h^{(0)}=t+v(x),\quad\\
&h^{(0)}\phi^{(0)}(x)=\epsilon^{(0)} \phi^{(0)}(x), 
\end{align}
where $x$, $\epsilon^{(0)}$, and $\phi^{(0)}$ are the 
relative coordinate, dimer energy, and dimer wave function respectively. 
For simplicity, the phase of $\phi^{(0)}$ is chosen to be real.

To discuss the effective interaction between two dimers, we consider 
a two-dimer system $Mm+Mm$ with two 
heavy $M$-particles placed at $x=-R/2$ on the left~(L) and $x=R/2$ on the right~(R) with separation  distance $R$.
The Hamiltonian for two light $m$-particles is written as 
\begin{align}\label{eq:hamil-two-dimer}
&H=h(1)+h(1'),\\
&h(i)=t(i)+U(x_i),\quad U(x)=v_\newlf(x)+v_\newrt(x),\\
&v_\textrm{\newlf}(x)=v(x+ R/2),\quad v_\textrm{\newrt}(x)=v(x- R/2),
\end{align}
where the first and second $m$ particles are labeled as $1$ and $1'$.
The energy $E(R)$ and the two-body wave function $\Psi(1,1')$ of the lowest 
state are given as 
\begin{align}
&E(R)=\epsilon_1+\epsilon_2, \\
&\Psi(1,1')={\cal A}\{\psi_1(1) \psi_2(1')\},
\end{align}
where $\epsilon_i$ and $\psi_i$ are the $i$th single-particle energy and state
obtained by solving the one-body problem of the single-particle Hamiltonian,  
$h(i)\psi_n(i)=\epsilon_n \psi_n(i)$. 
Because of the symmetry of the one-body potential $U(x)=U(-x)$, 
$\psi_n(i)$ are parity eigenstates with $\psi_1(x)=\psi_1(-x)$, $\psi_2(x)=-\psi_2(-x)$.
The effective dimer-dimer interaction is given by the relative energy $E(R)-2\epsilon^{(0)}$ measured 
from the two-dimer threshold energy.
This expression is exact for the heavy-mass limit, whereas it corresponds to 
the Born-Oppenheimer approximation for finite mass ratio.

\subsection{A cluster model for two-dimer system}
\subsubsection{Frozen dimer ansatz}
For a general discussion of the effective dimer-dimer interaction, 
we apply a cluster model to the two-dimer system with a frozen dimer ansatz
to approximately evaluate the energy $E(R)$. 
In this model, the system is expressed as the antisymmetrized product of ``atomic orbitals'' 
given by the isolated dimer wave functions around the left and right $M$-particles as 
\begin{align}
&\Phi(1,1')={\cal N}_0{\cal A}\{\phi^{(0)}_\newlf(1) \phi^{(0)}_\newrt(1')\},\\
&\phi^{(0)}_\textrm{\newlf}(i)=\phi^{(0)}(x_i+ R/2),\quad 
\phi^{(0)}_\textrm{\newrt}(i)=\phi^{(0)}(x_i- R/2),
\end{align}
where ${\cal N}_0$ is the normalization factor.
We introduce the following notation for the matrix elements
of one-body operators ${\cal O}$  with respect to 
$\phi^{(0)}_\textrm{\newlf}$ and $\phi^{(0)}_\textrm{\newrt}$ as 
\begin{align}
&\langle \phi^{(0)}_\newlf|{\cal O}|\phi^{(0)}_\newlf\rangle=
\langle {\cal O}\rangle_\textrm{\newll}, \quad \langle \phi^{(0)}_\newrt|{\cal O}|\phi^{(0)}_\newrt\rangle=
\langle {\cal O}\rangle_\textrm{\newrr}, \\
&\langle \phi^{(0)}_\newlf|{\cal O}|\phi^{(0)}_\newrt\rangle=
\langle {\cal O}\rangle_\textrm{\newlr}, \quad \langle \phi^{(0)}_\newrt|{\cal  O}|\phi^{(0)}_\newlf\rangle=
\langle {\cal O}\rangle_\textrm{\newrl}.
\end{align}
Here, the single-particle wave functions 
$\phi^{(0)}_\newlf$ and $\phi^{(0)}_\newrt$ are not orthogonal 
but has a nonzero norm overlap $\langle 1 \rangle_\textrm{\newlr}= \langle 1 \rangle_\textrm{\newrl}\ne 0$, 
which vanishes in the limit of large $R$. Nevertheless, 
the total wave function $\Phi(1,1')$ satisfies the 
Pauli principle (Fermi statistics) because of the antisymmetrizer, 
and ${\cal N}_0=1/\sqrt{1- \langle 1 \rangle_\textrm{\newrl}^2}$ is obtained from the 
normalization condition $\langle\Phi(1,1')|\Phi(1,1')\rangle=1$.

\subsubsection{Orthonormal bases sets: Molecular orbitals and orthonormal atomic orbitals}
\begin{figure*}[!htp]
\includegraphics[width=14 cm]{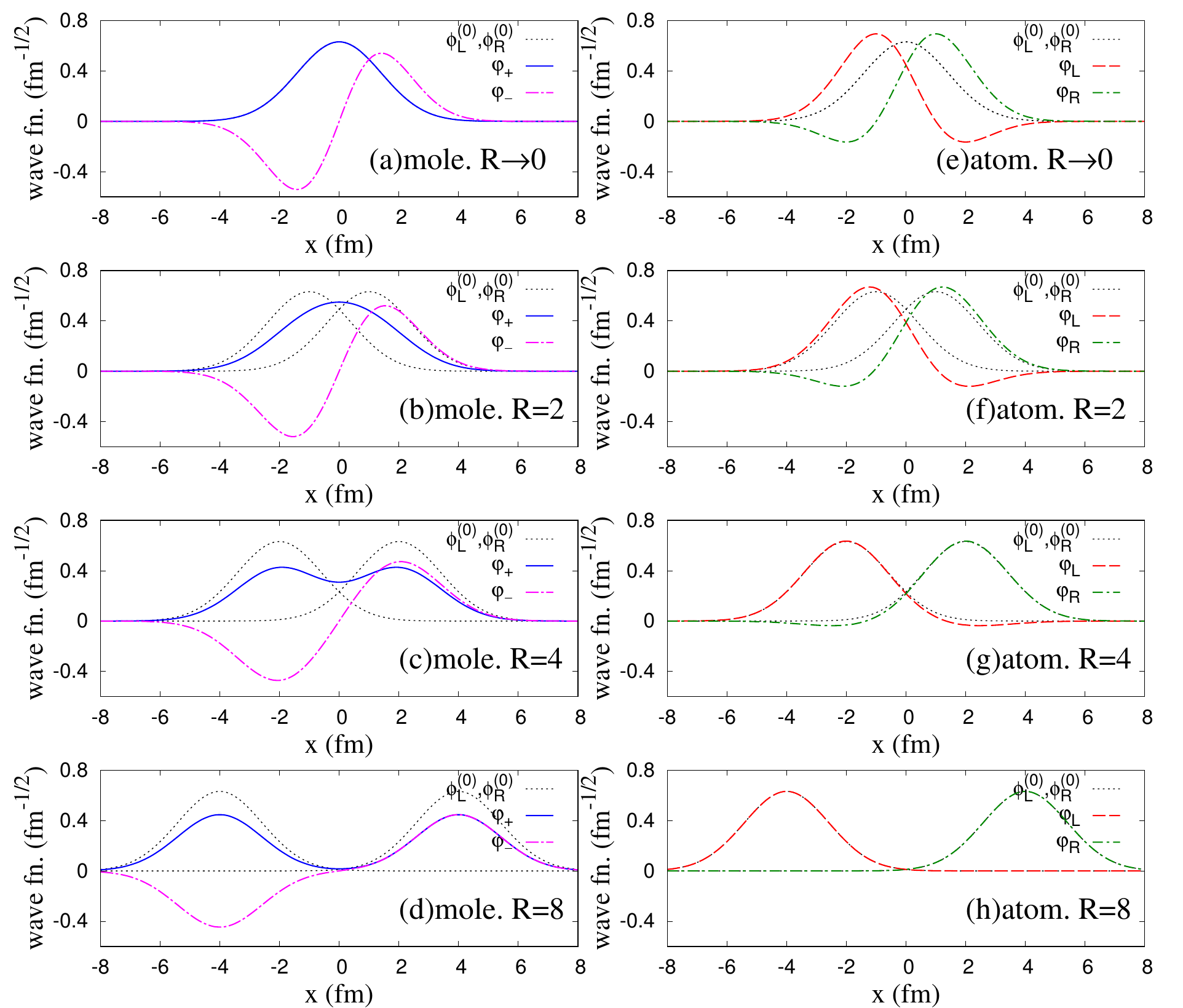}
\caption{(a)-(d) Molecular orbitals $\{\varphi_+,\varphi_-\}$ and (e)-(h) 
orthonormal atomic orbitals $\{\varphi_{\newlf},\varphi_{\newrt}\}$
of the two-dimer system for a Gaussian wave function of a single dimer 
$\phi^{(0)}(x)=\left(\frac{2\nu}{\pi}\right)^{1/4}e^{-\nu x^2}$
with $\nu=0.25$~fm$^{-2}$. The separation distances are chosen to be 
$R\to 0$, $R=2,4$, and 8 fm. 
The original atomic orbitals $\{\phi^{(0)}_{\newlf},\phi^{(0)}_{\newrt}\}$ on the left and right 
are also shown.
        \label{fig:mo-atom}}
\end{figure*}

The atomic orbitals $\phi^{(0)}_\textrm{\newlf}$ and $\phi^{(0)}_\textrm{\newrt}$ with small separation distance ($R$) overlap considerably with each other
and venture far into the Pauli forbidden region.
In this case it is more natural to view the total wave function $\Phi(1,1')$ rewritten using a new orthonormal basis set, 
taking into account the invariance of the normalized Slater determinant under any linear transformation of the basis vectors.
One choice is the basis set of ``molecular orbitals'' as
\begin{align}
&\Phi(1,1')={\cal A}\{\varphi_+(1) \varphi_-(1')\},\\
&\varphi_\pm (i)= \frac{1}{\sqrt{2 (1 \pm \langle 1 \rangle_\textrm{\newrl})}}
\{\phi^{(0)}_\newlf(i) \pm \phi^{(0)}_\newrt(i) \},
\end{align}
where $\varphi_+$ and $\varphi_-$ are positive- and negative-parity orbitals around whole system 
in analogy to covariant bonds of homonuclear diatomic molecules.
This expression with the molecular orbitals respects the parity symmetry of the one-body potential 
and is useful to discuss the two-dimer system in the overlapping region.
However, at long distances, 
the atomic orbital picture is more natural for the probability of an $m$-particle on the left or right. 
As yet another alternative basis set, ``orthonormal atomic orbitals'' can be also be defined as 
\begin{align}
\Phi(1,1')&={\cal A}\{\varphi_\newlf(1) \varphi_\newrt(1')\},\\
\varphi_\newlf(i)&=\frac{1}{\sqrt{2}}\left(\varphi_+(i) + \varphi_-(i)  \right),\\
\varphi_\newrt(i)&=\frac{1}{\sqrt{2}}\left(\varphi_+(i) - \varphi_-(i)  \right). 
\end{align}
It should be commented that the former set $\{\varphi_+, \varphi_-\}$
are obtained by solving the generalized eigenvalue problem for the $2\times 2$ matrices of the norm and Hamiltonian with respect to the basis states $\{\phi^{(0)}_\newlf,\phi^{(0)}_\newrt \}$. In contrast, the latter set of 
$\{\varphi_\newlf, \varphi_\newrt\}$ is obtained by solving the generalized eigenvalue problem for 
the norm and the position operator $x$. 

As a demonstration, we show the molecular orbitals $\{\varphi_+, \varphi_-\}$ and the orthonormal 
atomic orbitals $\{\varphi_\newlf, \varphi_\newrt\}$ for a Gaussian wave function
$\phi^{(0)}(x)=\left(\frac{2\nu}{\pi}\right)^{1/4}e^{-\nu x^2}$ with $\nu=0.25$~fm$^{-2}$
in Fig.~\ref{fig:mo-atom}. Fig.~\ref{fig:mo-atom}(a)-(d) compare  
the molecular orbitals with the original 
atomic orbitals for distances $R\to 0$, $R=2$, 4, and 8 fm. As the two dimers come close to each other, 
the positive-parity orbital is formed by merging the left and right atomic orbitals, 
while the negative-parity molecular orbital has an extra node at the origin. 
In Fig.~\ref{fig:mo-atom}(e)--(h), the orthonormal atomic orbitals 
$\{\varphi_\textrm{\newlf},\varphi_\textrm{\newrt}\}$ are compared with the 
original atomic orbitals $\{\phi^{(0)}_\newlf,\phi^{(0)}_\newrt \}$. 
At short distances $R\le 2$~fm, $\varphi_\textrm{\newlf}$ and $\varphi_\textrm{\newrt}$
are significantly distorted from the original orbitals 
because of antisymmetrization, while at long distances
the effect of antisymmetrization vanishes and 
they approach the original orbitals $\phi^{(0)}_\textrm{\newlf}$ and $\phi^{(0)}_\textrm{\newrt}$. 

\subsection{Effective interaction between two dimers}

Two dimers can not exist at the same position 
because of the Pauli principle between identical fermions.
This effect gives a repulsive contribution to the effective dimer-dimer interaction at short distance. 
As shown in Fig.~\ref{fig:mo-atom}(e)-(h), significant distortion occurs in 
the ``physical'' atomic orbitals, $\varphi_\textrm{\newlf}$ and $\varphi_\textrm{\newrt}$,
at short distances because of the antisymmetrization effect. 
As a result of the distortion, each dimer loses some internal energy. 
On the other hand, the $Mm$ potential term between different dimers can give an attractive contribution.
This shows that the effective dimer-dimer interaction is determined by 
the competition between the internal energy loss and the energy gain 
from the inter-cluster potential term. In this section, 
we investigate the two-dimer energy $E(R)$ with the frozen dimer ansatz
and discuss the effective dimer-dimer interaction. 

\subsubsection{Expression for general potentials}
The present model with the frozen dimer ansatz corresponds to an approximation of 
the single-particle wave functions $\psi$ in the two-dimer system 
with linear combination of the left and right atomic orbitals as
$\psi\approx \phi=c_\newlf\phi^{(0)}_\newlf+c_\newrt\phi^{(0)}_\newrt$, 
which is equivalent to a two-level problem given as
\begin{align}
\label{eq:hmatrix}&\braket{ H }_{\alpha\beta}=
\begin{pmatrix} 
\epsilon^{(0)}+\braket{v_\newrt}_{\newll} 
& \epsilon^{(0)}\braket{1}_{\newlr}+\braket{v_\newlf}_{\newlr} \\
 \epsilon^{(0)}\braket{1}_{\newrl}+ \braket{v_\newrt}_{\newrl} 
&\epsilon^{(0)}+\braket{v_\newlf}_{\newrr}
\end{pmatrix},\\
&\braket{\ 1\ }_{\alpha\beta}=
\begin{pmatrix} 
1&\braket{1}_{\newlr}\\
\braket{1}_{\newrl} & 1
\end{pmatrix},
\end{align}
with $(\alpha,\beta)=(\phi^{(0)}_\newlf,\phi^{(0)}_\newrt$).
By solving the generalized eigenvalue problem for these $2\times 2$ matrices, 
one obtains the molecular orbitals $\varphi_+$ and $\varphi_-$ as the eigensolutions 
with eigenvalues $\epsilon_+$ and $\epsilon_-$ as
\begin{align}
&\epsilon_\pm=\epsilon^{(0)}+\frac{\braket{v_\newrt}_{\newll}}{1\pm\braket{1}_{\newrl}}
\pm
\frac{\braket{v_\newrt}_{\newrl}}{1\pm\braket{1}_{\newrl}}.
\end{align}
The total energy of the two-dimer system and the relative energy measured from the threshold energy 
are obtained as 
\begin{align}
&E=\epsilon_+ + \epsilon_- \nonumber\\
&\quad =2\epsilon^{(0)} + \frac{2}{1-\braket{1}^2_{\newrl}}
\left(\braket{v_\newrt}_{\newll}
-\braket{1}_{\newrl} \braket{v_\newrt}_{\newrl} \right),\\
&\Delta E(R)\equiv E(R)-2\epsilon^{(0)}\nonumber\\
&\quad =\frac{2}{1-\braket{1}^2_{\newrl}}
\left(\braket{v_\newrt}_{\newll}
-\braket{1}_{\newrl} \braket{v_\newrt}_{\newrl} \right) \label{eq:dE-dir-ex}.
\end{align}
Note that the 
kinetic energy contribution does not explicitly appear in the present expression of $\Delta E(R)$, 
though it is implicitly contained in the exchange potential term with 
the relation $\braket{v_{\newrt}}_\newrl=\epsilon_0\braket{1}_\newrl-\braket{t}_\newrl$.

For the general case,
we consider an attractive potential $v(x)\le 0$ with a potential range
$r^{(0)}$ and a dimer wave function $\phi^{(0)}(x)\ge 0$ with 
a dimer size $b^{(0)}$.
Let us consider two terms in the expression $\braket{v_\newrt}_{\newll}-\braket{1}_{\newrl} \braket{v_\newrt}_{\newrl}$.
The first term, 
\begin{align}\label{eq:dir-pot}
\braket{v_\newrt}_{\newll}=\int v_\newrt(x) |\phi^{(0)}_\newlf(x)|^2 dx\le 0,
\end{align}
gives a negative~(attractive) contribution and is the direct potential term 
obtained by folding the right-side potential 
with the density $\rho^{(0)}_\newlf(x)\equiv |\phi^{(0)}_\newlf(x)|^2$ of the left-side 
atomic orbital. Roughly speaking, this term gives a finite contribution
in the $R < r^{(0)}+b^{(0)}$ region, where the dimer density has overlap with the 
closest edge of the external potential.

The second term,
\begin{align}\label{eq:ex-pot}
-\braket{1}_{\newrl} \braket{v_\newrt}_{\newrl}
=-\braket{1}_{\newrl} \int  \phi^{(0)}_\newlf(x) v_\newrt(x) \phi^{(0)}_\newrt(x) dx 
\ge 0, 
\end{align}
gives a positive~(repulsive) contribution corresponding to an exchange potential term. This term becomes significant in the 
$R< b^{(0)}+\min(b^{(0)},r^{(0)})$ region for the overlapping region of the two atomic orbitals and the right-side potential.

As an alternative expression, the sum of 
the direct and exchange potential terms can be rewritten as 
\begin{align} \label{eq:dE-ortho}
\braket{v_\newrt}_{\newll}
-\braket{1}_{\newrl} \braket{v_\newrt}_{\newrl}=
\langle \phi^{(0)}_\newlf |P^\perp_\newrt v_\newrt|\phi^{(0)}_\newlf\rangle,
\end{align}
where $P^\perp_\newrt\equiv 1-|\phi^{(0)}_\newrt \rangle\langle \phi^{(0)}_\newrt |$ is the projection operator onto the space orthogonal to 
$\phi^{(0)}_\newrt$. It means that 
the sum is the transition from $\phi^{(0)}_\newlf$ to the orthogonal 
component $P^\perp_\newrt|\phi^{(0)}_\newlf\rangle$ of the left-side particle by the external potential 
$v_\newrt$ on the right, and the exchange potential term arises from the orthogonal condition.


For the two-dimer energy $\Delta E(R)$ measured from the threshold in Eq.~\eqref{eq:dE-dir-ex},
the overall factor $\frac{2}{1-\braket{1}^2_{\newrl}}$ is positive because $|\braket{1}_{\newrl}|\le 1$. 
Therefore, the sign of the effective dimer-dimer interaction 
is determined by the competition between 
the attraction from the direct potential term of Eq.~\eqref{eq:dir-pot} 
and the repulsive effect from the exchange potential 
term of Eq.~\eqref{eq:ex-pot}. 

For the case where the local potential $v(x)$ 
has a range longer than the dimer size, $r^{(0)}> b^{(0)}$, 
the effective dimer-dimer interaction can be attractive in the intermediate distance region 
of $2b^{(0)} < R < r^{(0)}+b^{(0)}$. In this region the two atomic orbitals have almost 
no overlap $\phi^{(0)}_\newlf(x) \phi^{(0)}_\newrt(x)\sim 0$, and the  
exchange potential term is small compared with the direct potential term.
In the opposite case that $v(x)$ is a has a range shorter than the dimer size, $r^{(0)}< b^{(0)}$, the effective dimer-dimer interaction can be repulsive because of the
strong contribution from the exchange potential term. 
Also in the case of long-range 
but nonlocal potential $v(x,x')$, 
the effective dimer-dimer interaction may again be repulsive, 
because the nonlocality generally  
suppresses the matrix element $\braket{v_\newrt}_{\newll}$ in the direct potential term 
but enhances $\braket{v_\newrt}_{\newrl}$ in the exchange potential term.

All of the expressions derived in this section can be applied 
to dimer-dimer systems in three dimensions, just by replacing the one-dimensional integrals in 
the expectation values with three-dimensional integrals. We note that these
findings provide a conceptual foundation for the conclusions obtained
numerically in Ref.~\cite{Elhatisari:2016owd,Rokash:2016tqh}, that increasing
the range or strength of the local part of the particle-particle interaction
produces a more attractive cluster-cluster interaction.

\subsection{Effective dimer-dimer interaction with zero-range potential}
As an example of short-range potentials, we show that the effective dimer-dimer interactions 
with $M\gg m$ in 1D and 3D for a zero-range potential are always repulsive for any $R$. 
\subsubsection{Frozen cluster ansatz}
Firstly, we discuss the dimer-dimer interaction in 1D with the frozen  cluster ansatz.
For the delta potential 
\begin{align}
v(x)=-\frac{\hbar^2\kappa_0}{m}\delta(x), 
\end{align}
the energy and wave function of a single dimer are given as 
\begin{align}
\epsilon^{(0)}=-\frac{\hbar^2}{2m}\kappa_0^2,\quad
\phi^{(0)}(x)=\sqrt{\kappa_0}\textrm{e}^{-\kappa_0 |x|}, 
\end{align}
where $1/\kappa_0$ is roughly regarded as the dimer size $b^{(0)}$.
For the two-dimer system with the distance $R$, one can calculate 
matrix elements as
\begin{align}
&\braket{1}_{\newrl}= (1+\kappa_0 R)\textrm{e}^{-\kappa_0 R},\quad 
\braket{v_\newrt}_{\newrl}=2\epsilon^{(0)}  \textrm{e}^{-\kappa_0 R},\nonumber \\
&\braket{v_\newrt}_{\newll}=2\epsilon^{(0)} \textrm{e}^{-2\kappa_0 R},
\end{align}
and obtain energies for the positive- and negative-parity molecular orbitals
\begin{align}
&\epsilon_\pm=\epsilon^{(0)}+2\epsilon^{(0)}
\frac{\textrm{e}^{-2\kappa_0 R}\pm \textrm{e}^{-\kappa_0 R}}{1\pm \textrm{e}^{-\kappa_0 R}(1+\kappa_0 R)},
\end{align}
and the two-dimer energy from the threshold is
\begin{align}\label{eq:deltaE-app-1d}
\Delta E(R)=|\epsilon^{(0)}|\frac{4\kappa_0 R \textrm{e}^{-2\kappa_0 R}}{1-(1+\kappa_0 R)^2\textrm{e}^{-2\kappa_0 R}}>0.
\end{align}
This shows that the two dimers feel a repulsive dimer-dimer interaction for any $R$. 

Next, we show the result for the dimer-dimer interaction in 3D 
obtained with the frozen cluster ansatz.
For the renormalization of the single-delta potential in 3D, 
we assume that we have dimer with energy $\epsilon^{(0)}(< 0)$, corresponding with the bound state wave function
\begin{align}
\phi^{(0)}(\bvec{r)}=\sqrt{\frac{2\kappa_0}{4\pi}} \frac{\textrm{e}^{-\kappa_0 |\bvec{r}|}}{|\bvec{r}|},
\end{align}
with the definition $\kappa_0\equiv {\sqrt{2m |\epsilon^{(0)}|}}/{\hbar}$.
For the two-dimer system in 3D, we consider the single-particle energies  
for two delta potentials at $-{\bvec{R}}/{2}$~(on the left) and ${\bvec{R}}/{2}$~(on the right) with a distance $R=|\bvec{R}|.$ Using the frozen dimer ansatz, the matrix elements are obtained as
\begin{align}
&\braket{1}_{\newrl}= \textrm{e}^{-\kappa_0 R},\quad 
\braket{v_\newrt}_{\newrl}=2\epsilon^{(0)} \frac{ \textrm{e}^{-\kappa_0 R}}{\kappa_0 R},\nonumber \\
&\braket{v_\newrt}_{\newll}=0,
\end{align}
and the energies for the positive- and negative-parity molecular orbitals are
\begin{align}
&\epsilon_\pm=\epsilon^{(0)}\pm 2\epsilon^{(0)}
\frac{\textrm{e}^{-2\kappa_0 R}}{\kappa_0 R(1\pm \textrm{e}^{-\kappa_0 R})},
\end{align}
and the two-dimer energy measured from threshold is
\begin{align}\label{eq:deltaE-app-3d}
\Delta E(R)=|\epsilon^{(0)}|\frac{4  \textrm{e}^{-2\kappa_0 R}}{\kappa_0 R(1-\textrm{e}^{-2\kappa_0 R})
}>0,
\end{align}
indicating again a repulsive dimer-dimer interaction. 

Our results for the zero-range potential in 1D and 3D using the frozen dimer ansatz clearly 
show that the repulsive dimer-dimer interaction originates from 
the exchange potential term $\braket{v_\newrt}_{\newrl}$, {\it i.e.}, the antisymmetrization or Pauli  blocking effect. 

\subsubsection{Exact solution and asymptotic expansion}
We can also obtain exact solutions for the two-dimer energy in 1D and 3D by solving 
the two-delta potentials and see again the universal repulsion 
of the effective dimer-dimer interaction in the $M\gg m$ limit.

We express the exact energies $\epsilon^\textrm{exact}_\pm$ 
in terms of binding momenta $\kappa_\pm$ defined as 
\begin{align}
\epsilon^\textrm{exact}_\pm= -\frac{\hbar^2}{2m}\kappa^2_\pm.
\end{align}
For the exact solutions of the positive- and negative-parity bound states of the 1D two-delta potential, 
$\kappa_+$ and $\kappa_-$ are given 
as 
\begin{align}
&\kappa_+=\kappa_0\Bigl\{1+\frac{1}{\kappa_0R}W_0(\kappa_0R \textrm{e}^{-\kappa_0 R})\Bigr\}, \\
&\kappa_-=\kappa_0\Bigl\{1+\frac{1}{\kappa_0R}W_{-1}(\kappa_0R \textrm{e}^{-\kappa_0 R})\Bigr\}, 
\end{align}
where $W_0$ and $W_{-1}$ are branches of the 
Lambert $W$ function. With these solutions for $\kappa_\pm$,
the two-dimer energy measured from the threshold
is expressed as  
\begin{align}
\Delta E(R)= \epsilon^{(0)} \Bigl(\frac{\kappa^2_+}{\kappa^2_0}+ \frac{\kappa^2_-}{\kappa^2_0}-2 \Bigr ).
\end{align}
For large $\kappa_0 R$ we have the asymptotic forms 
\begin{align}
&\kappa_+\rightarrow \kappa_0 \Bigl(1+\textrm{e}^{-\kappa_0 R}  - \kappa_0 R \textrm{e}^{-2\kappa_0 R}+\cdots\Bigr ), \\
&\kappa_-\rightarrow \kappa_0 \Bigl(1-\textrm{e}^{-\kappa_0 R}  - \kappa_0 R \textrm{e}^{-2\kappa_0 R}+\cdots\Bigr ), 
\end{align}
and hence 
\begin{align}
\Delta E(R)= |\epsilon^{(0)}| \Bigl[ 4\kappa_0 R \textrm{e}^{-2\kappa_0 R} -2 \textrm{e}^{-2\kappa_0 R}+\cdots \Bigr ].
\end{align}
One can see that the leading term $|\epsilon^{(0)}|4\kappa_0 R  \textrm{e}^{-2\kappa_0 R}$ 
is consistent with that of the 
approximate result in Eq.~\eqref{eq:deltaE-app-1d} of the frozen dimer ansatz.

Similarly, the bound-state solutions for the 3D two-delta potential have binding momenta $\kappa_+$ and $\kappa_-$ of the form
 
\begin{align}\label{eq:kappaW-3D}
&\kappa_+=\kappa_0\Bigl\{1+\frac{1}{\kappa_0R}W_0(\textrm{e}^{-\kappa_0 R})\Bigr\}, \\
&\kappa_-=\kappa_0\Bigl\{1+\frac{1}{\kappa_0R}W_{-1}( \textrm{e}^{-\kappa_0 R})\Bigr\}.
\end{align}
The asymptotic forms for large $\kappa_0 R$ are
\begin{align}
&\kappa_+\rightarrow \kappa_0 \Bigl(1+\frac{\textrm{e}^{-\kappa_0 R}}{\kappa_0 R}  -
\frac{\textrm{e}^{-2\kappa_0 R}}{\kappa_0 R}+\cdots\Bigr ), \\
&\kappa_-\rightarrow 
\kappa_0 \Bigl(1-\frac{\textrm{e}^{-\kappa_0 R}}{\kappa_0 R}  -
\frac{\textrm{e}^{-2\kappa_0 R}}{\kappa_0 R}+\cdots\Bigr ),
\end{align}
and hence 
\begin{align}
\Delta E(R)= |\epsilon^{(0)}| \Bigl[ \frac{4 \textrm{e}^{-2\kappa_0 R}}{\kappa_0 R }
-\frac{2 \textrm{e}^{-2\kappa_0 R}}{(\kappa_0 R )^2}
+\cdots \Bigr ].
\end{align}
One can see again that the leading term 
is consistent with that in Eq.~\eqref{eq:deltaE-app-3d} of the frozen dimer ansatz.

\begin{figure}[!htp]
\includegraphics[width=7 cm]{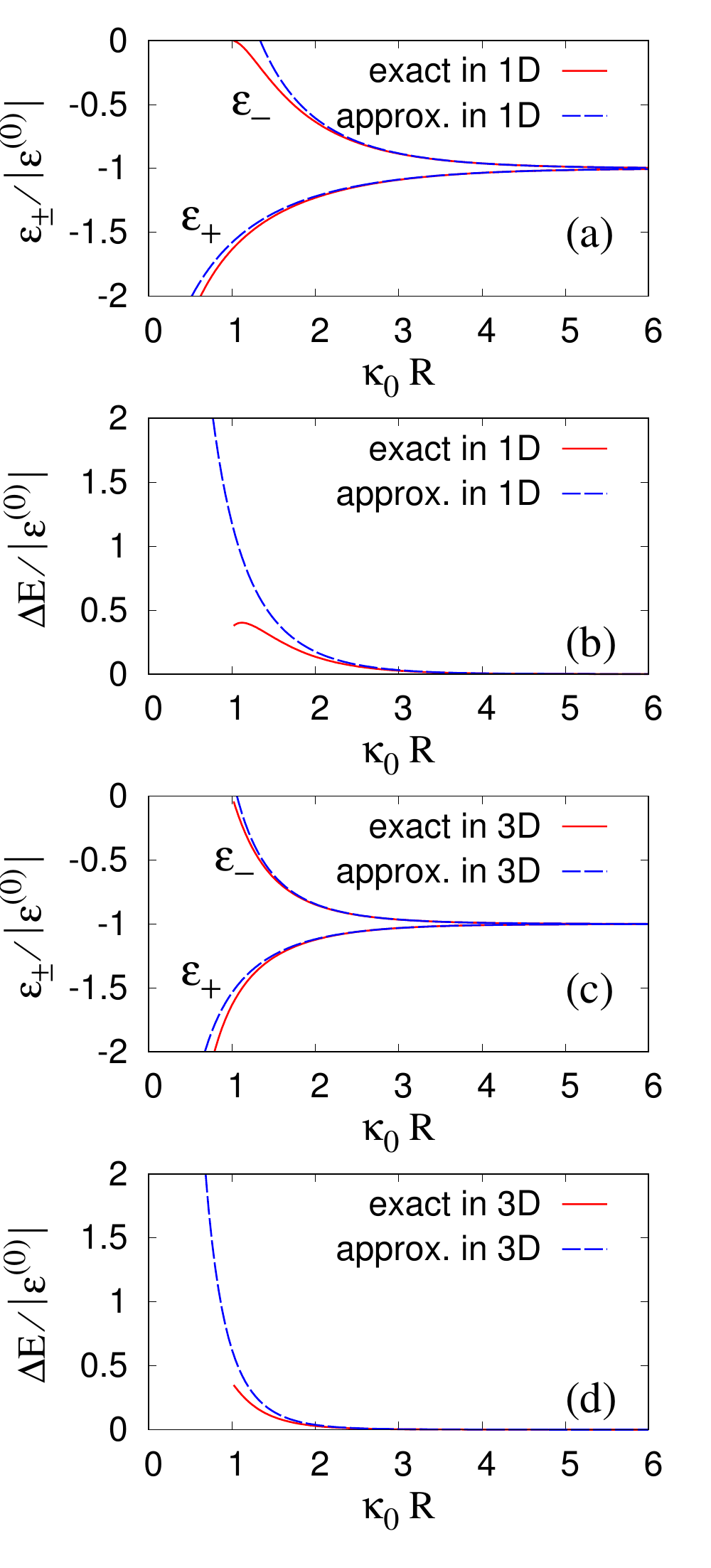}
\caption{Energies of the two-dimer system for delta potential
in the heavy-light ansatz in one dimension~(1D) and three dimensions~(3D).
The approximate values with the frozen dimer ansatz and exact values are compared. 
(a) Single-particle energies $\epsilon_\pm$ in 1D.
(b) The two-dimer energy
from the threshold energy, $\Delta E=\epsilon_++\epsilon_--2\epsilon^{(0)}$, in 1D.
(c) Single-particle energies $\epsilon_\pm$ in 3D.
(d) The two-dimer energy 
from the threshold energy in 3D. 
Energies are plotted in units of $1/|\epsilon^{(0)}|=2m/(\hbar^2 \kappa_0^2)$.
        \label{fig:ene-exact}}
\end{figure}

In Fig.~\ref{fig:ene-exact}, we compare the single-particle energies and the two-dimer energy
measured 
from the threshold energy
for exact solutions and approximate ones of the frozen cluster ansatz.
In the 1D results shown in Figs.~\ref{fig:ene-exact}(a) and (b), 
one can see that the frozen cluster ansatz is a good approximation for  
$ \kappa_0 R\gtrsim 2$,  
but gets worse for $ R \lesssim 2/\kappa_0$, 
where two dimers are closer than twice of the dimer size $b^{(0)}\sim 1/\kappa_0$.
For the 3D case, it is a good approximation for $ \kappa_0 R\gtrsim 1.5$ as shown in Figs.~\ref{fig:ene-exact}(c) and (d).  
The detailed solution for the single-particle energies and wave functions 
in the 1D two-delta potential are described in Appendix \ref{app:two-delta}.

\section{Nuclear systems of two clusters: $d+d$, $t+t$, and $\alpha+\alpha$}
\label{sec:two-cluster}

\subsection{Cluster model wave functions}
We now discuss the effective interactions between two nuclear clusters by applying the Brink-Bloch cluster model \cite{brink66}.  We consider $d+d$, $t+t$, and $\alpha+\alpha$ systems with
$d$, $t$, and $\alpha$ clusters consisting of two, three and four 
nucleons, respectively. 
We denote the mass number of a cluster as $c$ $(c=2,3,4)$ and use a label
``$c+c$'' for the two-cluster systems. 

\subsubsection{Single-cluster wave function}
In the cluster model, a single 
cluster is assumed to be a $c$-nucleon state with the harmonic oscillator 
$0s$-orbit configuration noted as $(0s)^c$.
The wave function for the cluster placed at $\bvec{R}_1$
is written as a product of single-particle Gaussian wave functions as 
\begin{align}
&\Phi^{c}_{\bvec{R}_1}(1,\ldots,c) \nonumber\\
&= {\cal{A}} \Bigl\{\phi^{(0)}_{\bvec{R}_1}({\bvec{r}}_1)\cdots
\phi^{(0)}_{\bvec{R}_1}({\bvec{r}}_c)\otimes\chi_c(s_1,\ldots,s_c)\Bigr\},\\
&\phi^{(0)}_{\bvec{R_1}}({\bvec{r}})=
 \left(\frac{2\nu}{\pi}\right)^{3/4}
\exp\bigl[-\nu({\bvec{r}}-\bvec{R}_1)^2\bigr],
\end{align}
where $s_i$ indicates the nucleon spin and isospin degrees of freedom of the $i$th nucleon,
and $\chi_c$ is the spin and isospin function of the $(S=1,T=0)$, $(S=1/2,T=1/2)$, and $(S=0,T=0)$ 
states for the deuteron, triton, and $\alpha$ clusters, respectively. 

\subsubsection{Two-cluster wave function}
For the $d$-cluster with $S=1$, we consider the spin-aligned 
$[d+d]_{S=2}$ and spin-opposed $[d+d]_{S=0}$ states.
Similarly, for the $t$-cluster with $S=1/2$, 
the spin-aligned $[t+t]_{S=1}$ and spin-opposed $[t+t]_{S=0}$ states are considered.

The wave function of a two-cluster system with separation distance $R$
is given as 
\begin{align}
&\Phi_{c+c}(\bvec{R}; 1,\ldots,c,1',\dots,c') \nonumber\\
&= ({\cal N}_0)^{n_\textrm{id}}{\cal{A}} \Bigl\{
\Phi^{c}_{-\frac{\bvec{R}}{2}}(1,\ldots,c)\Phi^{c}_{\frac{\bvec{R}}{2}}(1',\ldots,c')\Bigr\}
\nonumber\\
&= ({\cal N}_0)^{n_\textrm{id}} {\cal{A}} \Bigl\{
\phi^{(0)}_{-\frac{\bvec{R}}{2}}({\bvec{r}}_1)\cdots
\phi^{(0)}_{-\frac{\bvec{R}}{2}}({\bvec{r}}_c)\phi^{(0)}_{\frac{\bvec{R}}{2}}({\bvec{r}}_{1'})\cdots
\phi^{(0)}_{\frac{\bvec{R}}{2}}({\bvec{r}}_{c'})
\nonumber\\
& \otimes\bigl[\chi_c(s_1,\ldots,s_c)\chi_c(s_{1'},\ldots,s_{c'})\bigr]_{S} 
\Bigr\},\label{eq:ccatomic}
\end{align}
where spins of two clusters are coupled to $S$ in total, and 
$\bvec{R}$ is chosen to be $(0,0,R)$ on the $z$ axis.
$n_\textrm{id}$ is the number of pairs of identical nucleons. $n_\textrm{id}=2,3,4$ for $[d+d]_{S=2},[t+t]_{S=1},[\alpha+\alpha]_{S=0}$ respectively, and $n_\textrm{id}=0,2$ for $[d+d]_{S=0},[t+t]_{S=0}$ respectively. 

\begin{figure}[!htp]
\includegraphics[width=8.6 cm]{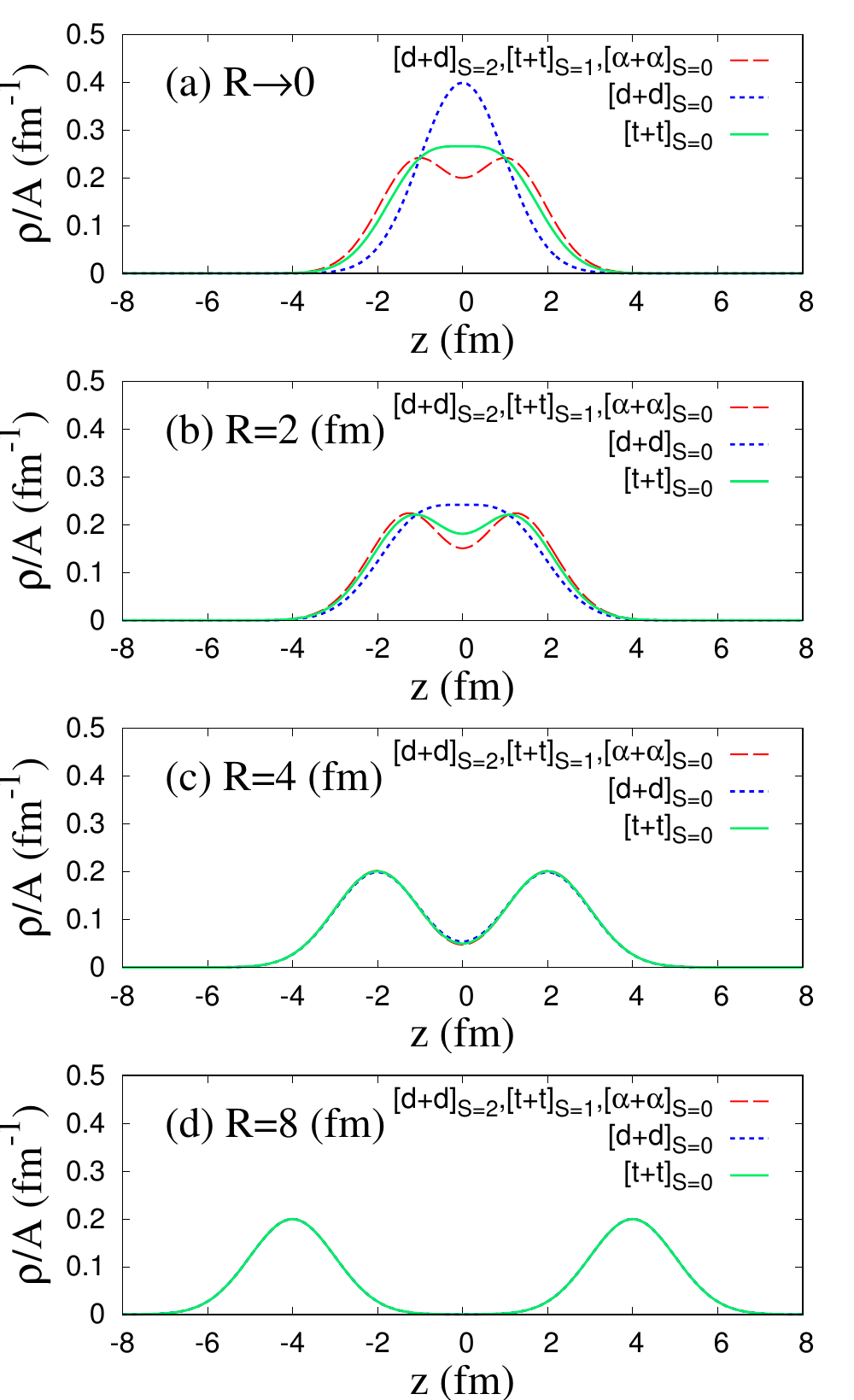}
\caption{
Nuclear matter density of the two-cluster wave functions for $d+d$, $t+t$, and $\alpha+\alpha$ systems
with distances $R\to 0$, $R=2,4$, and 8 fm.
The densities are  
integrated over $x$ and $y$ and normalized with the mass number $A=2c$ as $\rho(z)/A$. 
The normalized densities of the $[d+d]_{S=2}$, $[t+t]_{S=1}$, and $[\alpha+\alpha]_{S=0}$ states, which are
consistent with each other, are plotted with dashed lines, and those of $[d+d]_{S=0}$ 
and $[t+t]_{S=0}$ states are shown by dotted and solid lines, respectively. 
        \label{fig:dens}}
\end{figure}

The nuclear matter densities of two-cluster wave functions are shown in Fig.~\ref{fig:dens}.
The $[d+d]_{S=2}$, $[t+t]_{S=1}$, and $[\alpha+\alpha]_{S=0}$ systems are composed of 
$d=(p_\uparrow n_\uparrow)$, $t=(p_\uparrow n_\uparrow n_\downarrow)$, and 
$\alpha=(p_\uparrow p_\downarrow n_\uparrow n_\downarrow)$, respectively, 
and they show a dumbbell-like drop in the density in the $R\lesssim 2$ fm region, indicating the strong Pauli blocking effects of the identical-nucleon pairs.
On the other hand, in the $[d+d]_{S=0}$ system with no identical-nucleon pairs,
the two clusters can penetrate each other without any Pauli blocking and merge into a $^4$He state 
with an $(0s)^4$ configuration in the $R\to 0$ limit. 
The $[t+t]_{S=0}$ state containing two identical-nucleon pairs shows a 
weaker Pauli blocking effect than the $[t+t]_{S=1}$ state.

For the $[d+d]_{S=2}$, $[t+t]_{S=1}$, and $[\alpha+\alpha]_{S=0}$ systems, 
the total wave function $\Phi_{c+c}(\bvec{R})$ is expressed by 
a Slater determinant of non-orthonormal atomic orbitals
$\{\phi^{(0)}_{-\frac{\bvec{R}}{2}}(i),\phi^{(0)}_{\frac{\bvec{R}}{2}}(i')\}$, which can be 
transformed into the molecular orbitals set $\{\varphi_+(i),\varphi_-(i') \}$ 
or the orthonormal atomic orbitals set $\{\varphi_\newlf(i),\varphi_\newrt(i') \}$ 
under invariance of the total wave function 
as described previously in Section \ref{sec:heavy-light}.

\subsubsection{Parity and orbital-angular-momentum projections}
We consider the parity $(\pi)$ and orbital-angular-momentum ($L$) projection of 
the two-cluster wave functions as
\begin{align}
&\Phi^\pi_{c+c}(R)=P^\pi\Phi_{c+c}(\bvec{R}),\\
&\Phi^{L\pi}_{c+c}(R)=P^L P^\pi\Phi_{c+c}(\bvec{R}),
\end{align}
with the $L$ and $\pi$ projection operators $P^L$ and $P^\pi$. 
The intrinsic energy $E_\textrm{int}(R)$ at a distance $R$ is calculated 
using the $\pi$-projected wave function without the $L$-projection as 
\begin{align}
&E^\textrm{int}_{c+c}(R)=\frac{\langle \Phi^\pi_{c+c}(R)|H|\Phi^\pi_{c+c}(R)\rangle}
{\langle \Phi^\pi_{c+c}(R)|\Phi^\pi_{c+c}(R)\rangle}.
\end{align}
Similarly the  $L^\pi$-projected energy is calculated 
with the $L^\pi$-projected wave function as 
\begin{align}
&E^{L\pi}_{c+c}(R)=\frac{\langle \Phi^{L\pi}_{c+c}(R)|H|\Phi^{L\pi}_{c+c}(R)\rangle}
{\langle \Phi^{L\pi}_{c+c}(R)|\Phi^{L\pi}_{c+c}(R)\rangle}.
\end{align}
We take $\pi=-$ and $L=1$ ($P$-wave) for the $[t+t]_{S=1}$ system as required by antisymmetry, and $\pi=+$ and $L=0$ ($S$-wave) for the other systems.

The total angular momentum and parity are $J^\pi=0^-,1^-,2^-$ for the $[t+t]_{S=1}$ system, and 
$J^\pi=S^\pi$ for the other systems. 
Strictly speaking, the
$J^\pi=2^+$ and $0^+$ states are coupled in the $d+d$ system and 
the $J^\pi=0^-,1^-$ and $2^-$ states are coupled in the $[t+t]_{S=1}$ system 
because of the $NN$ spin-orbit and tensor interactions, but we omit such the channel couplings 
due to our assumption of effective $NN$ central forces 
for simplicity.  

\subsubsection{GCM calculation of two-cluster systems}
We calculate the energy $E_{c+c}$ of the ground states of two-cluster systems
with the GCM \cite{GCM1,GCM2}
by superposing $L^\pi$-projected wave functions
\begin{eqnarray}\label{eq:gcm-BB}
\Psi^{\textrm{GCM}}_{c+c}&=&\sum_k c_k  \Phi^{L\pi}_{c+c} (R_k),
\end{eqnarray}
where coefficients $c_k$ are determined by 
solving the discretized Hill-Wheeler equation~\cite{GCM1}, {\it i.e.}, solving 
the generalized eigenvalue problem for norm and Hamiltonian matrices with respect to $k$.
This GCM calculation corresponds to
optimization of the inter-cluster wave function as described in 
Appendix~\ref{app:relative-wf}.

We also perform one-dimensional GCM calculations (1d-GCM) by superposing the $\pi$-projected wave functions
instead of the $L^\pi$ projected ones as 
\begin{eqnarray}\label{eq:1d-gcm-BB}
\Psi^{\textrm{1d-GCM}}_{c+c} &=&\sum_k c_k  \Phi^{\pm}_{c+c} (R_k).
\end{eqnarray}
In the 1d-GCM calculation, 
all nucleons ($i=1,\ldots,c,1'\ldots,c'$) are confined 
for the $x$ and $y$ directions
in the same Gaussian orbit 
$\left(\frac{2\nu}{\pi}\right)^{1/2}\exp\bigl[-\nu(x_i^2+y_i^2)]$, 
whereas the inter-cluster motion in the $z$ direction is optimized 
by the superposition. After diagonalization of the norm and Hamiltonian matrices, 
one obtains the 1d-GCM energy $E^\textrm{1-dim}$ for
the lowest solution of the one-dimensional motion. 

\subsection{Hamiltonian and effective nuclear force}
The Hamiltonian of nuclear systems is given as 
\begin{align}
&H=\sum_i t_i -T_\textrm{cm} +\sum_{i<j}v_{N}(i,j),\\
&t_i=-\frac{\hbar^2}{2M_N}\frac{\partial^2}{\partial \bvec{r}_i^2},
\quad T_\textrm{cm}=-\frac{\hbar^2}{2AM_N}\frac{\partial^2}{\partial \bvec{r}_\textrm{cm}^2},
\end{align}
where $M_N$ is the nucleon mass and $v_{N}$ is the effective two-body nuclear force. 
In the cluster model, the center of mass (cm) motion 
can be separated and the cm kinetic energy term $T_\textrm{cm}$ is constant, 
$T_\textrm{cm}=(3/4)\hbar\omega$ with $\omega= 2\hbar^2\nu/M_N$.

As for the effective $NN$ force, we use a finite-range central force 
of the Volkov No.2 force \cite{VOLKOV}, which can be written
with the triple-even~($^3E$), singlet-even~($^1E$), triplet-odd~($^3O$), 
and singlet-odd~($^1O$) terms as 
\begin{align}
&v_{N}(1,2)=V_{N}(r)\times\bigl[f_{3E}P(^3E)+ f_{1E}P(^1E)\nonumber\\
&\qquad +f_{3O}P(^3O)+f_{1O}P(^1O)\bigr],\label{eq:volkov-f}\\
&r\equiv |\bvec{r}_2-\bvec{r}_1|,
\end{align}
where the radial function $V_{N}(r)$ is given by a two-range Gaussian form.
In the original expression, the Volkov force is given by the Wigner, Bertlett, Heisenberg, and 
Majorana terms. Details of parametrization of the Volkov No.2 force and its relation to 
ratios $f_{3E}$, $f_{1E}$, $f_{3O}$, and $f_{1O}$ in Eq.~\eqref{eq:volkov-f}
are explained in Appendix \ref{app:volkov}. 

\begin{table*}[!ht]
\caption{Parameter sets of the Volkov No.2 force
for four-types of the $NN$ forces, $v^\textrm{SU4}_{N}$~(SU4-even), $v^\textrm{tuned}_{N}$~(tuned), 
$v^\textrm{st-ind}_{N}$~(state-independent), and $3v^\textrm{SU4}_{N}$~(strong-even) forces.
Details of the strength parameters ($f_{3E}$, $f_{1E}$, $f_{3O}$, and $f_{1O}$) 
for the $^3E$, $^1E$, $^3O$, and $^1O$ terms and the parameters ($W$ $B$, $H$, and $M$) 
for the Wigner, Bartlett, Heisenberg, and Majorana 
terms are described in Appendix \ref{app:volkov}. 
 \label{tab:volkov}
}
\begin{center}
\begin{tabular}{c|cccc|cccccc}
\hline
                &       $f_{3E}$        &$f_{1E}$       &$f_{3O}$       &$f_{1O}$       &       $W$     &       $B$     &       $H$     &       $M$     \\
$v^\textrm{SU4}_{N}$ SU4-even   &       1       &       1       &       0       &       0 &       0.5     &       0       &       0       &       0.5     \\
$v^\textrm{tuned}_{N}$ tuned    &       1.3     &       0.7     &       $-0.2$  &       $-0.2$  &       0.4     &       0.15    &       0.15    &       0.6     \\
$v^\textrm{st-ind}_{N}$ state-independent &     1.3     &       1.3     &       1.3     &       1.3      &       1       &       0.3     &       0       &       0\\
$3v^\textrm{SU4}_{N}$ strong-even &     3       &       3       &       0       &       0 &       1.5     &       0       &       0       &       1.5             \\
\hline
\end{tabular}
\end{center}
\end{table*}

The parameter sets \{$f_{3E},f_{1E},f_{3O},f_{1O}$\} 
adopted in the present calculation are summarized in Table \ref{tab:volkov}.
The first set is a purely even-parity force with SU4 symmetry as 
\begin{align}
v^\textrm{SU4}_{N}&=V_{N}(r)\bigl[P(^3E)+P(^1E) \bigr],
\end{align}
which we call the SU4-even force. This force 
acts on spatial even components of otherwise-nucleon pairs, 
$(p_\uparrow p_\downarrow)$, $(p_\uparrow n_\uparrow)$, 
$(p_\uparrow n_\downarrow)$, and $(n_\uparrow n_\downarrow)$ with the same strength.
Note that 
the $[d+d]_{S=2}$ state is equivalent 
to a four-neutron system of two dineutrons $(nn)+(nn)$ in the case of SU4-symmetric forces.
The second set is a tuned force 
\begin{align}
v^\textrm{tuned}_{N}&=V_{N}(r)\bigl[1.3P(^3E)+0.7P(^1E)\nonumber\\
& -0.2P(^3O)-0.2P(^1O) \bigr],
\end{align}
adjusted to fit the experimental data of 
$S$-wave $NN$ scattering lengths in the spin-triplet and spin-singlet channels and 
$\alpha$-$\alpha$ scattering phase shifts. 
This tuned force contains 
a stronger $^3E$ force and a weaker $^1E$ force with the ratio of 1.3/0.7 
to describe a bound deuteron state and an unbound $nn$ state. 

In addition, we consider two optional sets to make 
the $[d+d]_{S=2}$ system to be bound, 
which do not describe physical nuclear systems. 
One is a strong-even $NN$ force
\begin{align}
3v^\textrm{SU4}_{N}=V_{N}(r)\bigl[3P(^3E)+3P(^1E) \bigr],
\end{align}
which is three times as strong as the SU4-even $NN$ force.
The other is a state-independent $NN$ force containing $^1E$, $^3O$, and $^1O$ attraction 
with the same strength as the $^3E$ component of $v^\textrm{tuned}_{N}$ 
\begin{align}
v^\textrm{st-ind}_{N}&=V_{N}(r)\bigl[1.3 P(^3E)+1.3 P(^1E)\nonumber\\
& + 1.3 P(^3O)+ 1.3 P(^1O) \bigr].
\end{align}

\subsection{Energy of single-cluster systems}

\begin{figure}[!htp]
\includegraphics[width=7 cm]{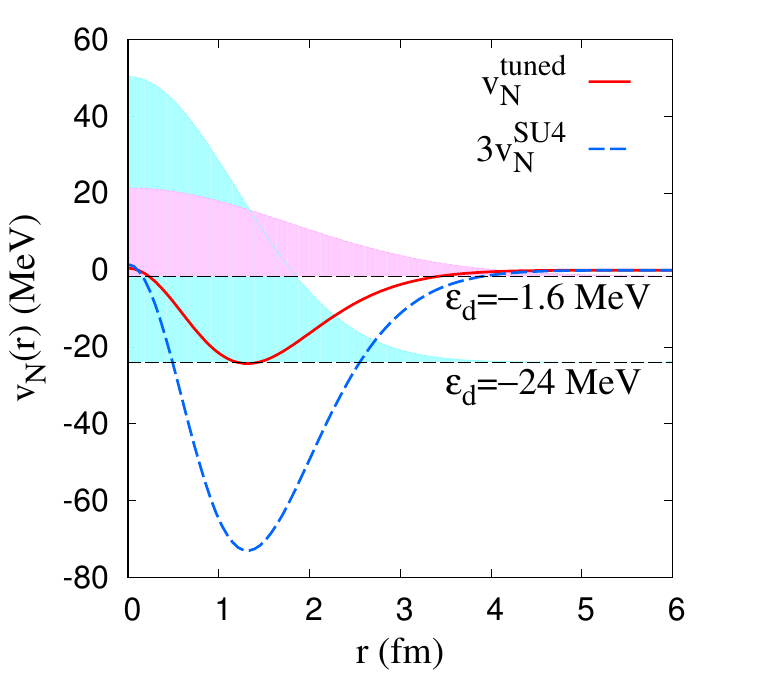}
\caption{
Radial dependence of the $^3E$ component of the tuned ($v^\textrm{tuned}_{N}$) and 
strong-even ($3v^\textrm{SU4}_{N}$) forces.
The internal wave function $\Phi^d(r)\propto \exp[-2\nu r^2]$
of a $d$-cluster with the $(0s)^2$ configuration with 
$\nu=0.16$ fm$^{-2}$ for the tuned force
and that with $\nu=0.35$ fm$^{-2}$ for the strong-even force are shown 
by pink and light-blue colored areas, respectively, in an arbitrary unit.
        \label{fig:pot}}
\end{figure}

In Table~\ref{tab:two-cluster}, we show the total, kinetic and potential energies 
for a single-cluster system of $d$, $t$, and $\alpha$ calculated with the $(0s)^c$ configurations.
Values of the width parameter $\nu$ used in the present calculation 
are also listed in the table.
For the $v^\textrm{SU4}_{N}$ force, $\nu$ is fixed to be 
a common value $\nu=0.25$ fm$^{-2}$, which reproduces the root-mean-square~(rms) radius of an $\alpha$ particle.
For other three forces, $v^\textrm{tuned}_{N}$,  $v^\textrm{st-ind}_{N}$, and $3v^\textrm{SU4}_{N}$,
we use the values $\nu=0.16$,  $0.16$, and $0.35$ fm$^{-2}$,  respectively, 
which are optimized to minimize the $d$-cluster energy.

Let us compare the energies of the $d$, $t$, and $\alpha$ clusters obtained with 
the SU4-even force ($v^\textrm{SU4}_{N}$).
As the constituent nucleons increases,  
the single-cluster system obtains a deeper binding
because the kinetic energy loss is proportional to $c-1$ while
the potential energy gain is proportional to the number $c(c-1)/2$ of nucleon
pairs.
The tuned $NN$ force ($v^\textrm{tuned}_{N}$)
gives a bound $d$ state at the energy $\epsilon_d=-1.6$~MeV for $\nu=0.16$ fm$^{-2}$, while 
it gives the same $\alpha$ energy $\epsilon_\alpha=-28.7$~MeV  
as the SU4-even force. 
The state-independent force $v^\textrm{st-ind}_{N}$ obtains 
$\epsilon_d=-1.6$~MeV, same as the tuned $NN$ force because  
the $NN$ force in the $^3E$ channel is unchanged. 

The strong-even $NN$ force~($3v^\textrm{SU4}_{N}$) gives a deeply bound $d$ state 
with $\nu=0.35$ fm$^{-2}$ at $\epsilon_d=-24.0$~MeV. 
The radial dependence of the $^3E$ component
and the deuteron wave function for the tuned and strong-even $NN$ forces 
are shown in Fig.~\ref{fig:pot}. Compared with the tuned force, the $d$-cluster  for the strong-even force is more deeply bound and 
the cluster size is much smaller.

\begin{table}[!ht]
\caption{Energies of single-cluster and two-cluster systems calculated with the cluster model
using four types of the $NN$ force. 
For single-cluster systems,  
the total~($\epsilon_c$), kinetic~($T$), and potential~($V$) energies 
are shown together with the adopted $\nu$ values (fm$^{-2}$).
For two-cluster systems, GCM energies measured from the $c+c$ threshold energy $(2\epsilon_c)$
and 1d-GCM energies relative to 
the one-dimensional $c+c$ decay threshold energy $(2\epsilon_c+\hbar\omega/2)$ are shown. 
For the $[t+t]_{S=1}$ system, the GCM result for the $L^\pi=1^-$ state and 
the 1d-GCM result for the $\pi=-$ state are shown. For other systems, 
the GCM result for the $L^\pi=0^+$ state and the 1d-GCM result for the $\pi=+$ state
are shown. For unbound systems, 
positive energies are obtained in the present framework of a 
bound state approximation with $R\le 10$ fm. 
The energy unit is MeV. 
 \label{tab:two-cluster}
}
\begin{center}
\begin{tabular}{ccccc|cccccc}
\hline
\multicolumn{8}{l}{$v^\textrm{SU4}_{N}$: SU4}\\
        &       $\nu$   &       $\epsilon_c$    &       $T$     &       $V$     &               &       $\Delta E_{c+c}$        &       $\Delta E^\textrm{1-dim}_{c+c}$ \\
$d$     &       0.25    &$      3.0     $&      15.6    &$      -12.6   $&      $[d+d]_{S=2}$   &       unbd.(1.34)     &       unbd.(0.98)     \\
        &               &$              $&              &$              $&      $[d+d]_{S=0}$   &$      -34.8         $&$     -45.1   $\\
$t$     &       0.25    &$      -6.6    $&      31.1    &$      -37.7   $&      $[t+t]_{S=1}$   &       unbd.(1.14)     &$      -1.12         $\\
        &               &$              $&              &$              $&      $[t+t]_{S=0}$   &$      -12.8         $&$     -21.5   $\\
$\alpha$        &       0.25    &$      -28.7   $&      46.7    &$      -75.3         $&      $[\alpha+\alpha]_{S=0}$ &$      -7.6    $&$     -12.5   $\\
\ \\
\multicolumn{8}{l}{$v^\textrm{tuned}_{N}$: tuned}\\             
        &       $\nu$   &       $\epsilon_c$    &       $T$     &       $V$     &               &       $\Delta E_{c+c}$        &       $\Delta E^\textrm{1-dim}_{c+c}$ \\
$d$     &       0.16    &$      -1.6    $&      10.0    &$      -11.6   $&      $[d+d]_{S=2}$   &       unbd.(2.9)      &       unbd.(0.97)     \\
$\alpha$        &       0.25    &$      -28.7   $&      46.7    &$      -75.3         $&      $[t+t]_{S=1}$   &$      -2.7    $&$     -3.6    $\\
\ \\
\multicolumn{8}{l}{$3v^\textrm{SU4}_{N}$: strong even}\\                                                                                                                                
        &       $\nu$   &       $\epsilon_c$    &       $T$     &       $V$     &               &       $\Delta E_{c+c}$        &       $\Delta E^\textrm{1-dim}_{c+c}$ \\
$d$     &       0.35    &$      -24.0   $&      21.8    &$      -45.7   $&      $[d+d]_{S=2}$   &$      -0.36         $&$     -5.1    $\\
\ \\
\multicolumn{8}{l}{$v^\textrm{st-ind}_{N}$: state-independent}\\           
        &       $\nu$   &       $\epsilon_c$    &       $T$     &       $V$     &               &       $\Delta E_{c+c}$        &       $\Delta E^\textrm{1-dim}_{c+c}$ \\
$d$     &       0.16    &$      -1.6    $&      10.0    &$      -11.6   $&      $[d+d]_{S=2}$   &$      -1.1         $&$     -6.1 $  \\
\hline
\end{tabular}
\end{center}
\end{table}

\subsection{Two-cluster systems}

\subsubsection{GCM and 1d-GCM resuts}
To obtain the lowest states of two-cluster systems, 
we perform the GCM calculations 
using the two-cluster wave functions with $R_k=0.5,~1,\ldots,10$ fm. 
The calculations correspond to a bound state approximation in a finite box boundary $R_k\le 10$ fm. 
We also perform the 1d-GCM calculations to check whether two clusters 
effectively feel an attraction forming a one-dimension bound state or not. 

In Table~\ref{tab:two-cluster}, the GCM and 1d-GCM energies of two-cluster systems 
are listed. For the GCM result, the energy is measured from 
the $c+c$ threshold energy as 
$\Delta E_{c+c}\equiv E_{c+c}-2\epsilon_c$. For the 1d-GCM result, 
the energy is measured from the one-dimensional $c+c$ decay threshold 
\begin{align}
\Delta E^\textrm{1-dim}_{c+c}=E^\textrm{1-dim}_{c+c}-(2\epsilon_c+\frac{1}{2}\hbar\omega)
\end{align}
are shown. Here the one-dimensional decay threshold contains an extra 
kinetic energy cost $2(\hbar\omega/4)$ for localization in two directions on the $xy$ plane.

For the $[d+d]_{S=2}$ system with the $v^\textrm{tuned}_{N}$(tuned) and $v^\textrm{SU4}_{N}$ (SU4-even) forces, 
no bound state is obtained in both the GCM and 1d-GCM calculations, indicating that the effective 
interaction  between to $d$-clusters in the $S=2$ channel is repulsive. For the 
$[\alpha+\alpha]_{S=0}$ system with the $v^\textrm{tuned}_{N}$(tuned) force, 
the GCM calculation obtains a weakly bound state without the Coulomb force, 
but an unbound state with the Coulomb force, consistent with the observed quasi-bound 
$2\alpha$ state of $^8$Be($0^+$).
The $[t+t]_{S=1}$ system with the $v^\textrm{SU4}_{N}$ (SU4-even) force,
is not bound in the GCM calculation but bound in the 1d-GCM calculation
meaning that the effective interaction between two $t$-clusters in the $S=1$ channel 
is a weak attraction.
 
In the $[d+d]_{S=0}$ system for the spin-opposed $d+d$ in the $S=0$ channel, 
two $d$-clusters are deeply bound 
and form an $\alpha$ particle because there is no Pauli blocking in this system.
Also the $[t+t]_{S=0}$ system forms a bound state because of a weaker Pauli effect
than the $[t+t]_{S=1}$ system.

\subsubsection{Energy curves of two-cluster systems}
\begin{figure*}[!htp]
\includegraphics[width=16 cm]{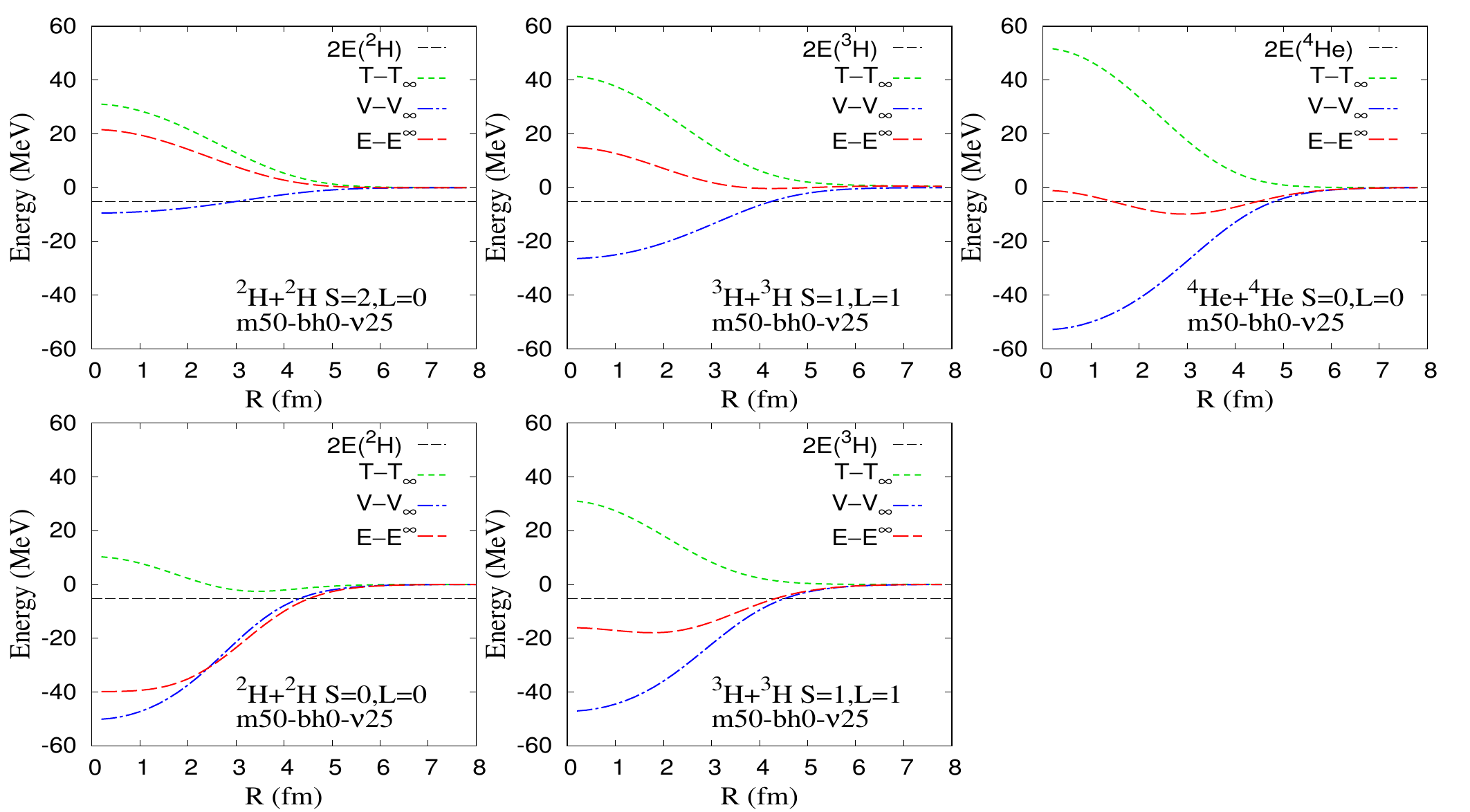}
\caption{$L^\pi$-projected energies $E^{L\pi}_{c+c}(R)$ of two-cluster systems with
separation distance $R$ calculated with the SU4-even $NN$ force~($v^\textrm{SU4}_{N}$).
Total $(E)$, kinetic ($T$), and potential $(V)$ energy contributions of  
(a) $[d+d]_{S=2}$, (b) $[\alpha+\alpha]_{S=0}$, (d) $[d+d]_{S=0}$, and (e) $[t+t]_{S=0}$
for the $L^\pi=0^+$ states and those of (c) $[t+t]_{S=1}$ for the $L^\pi=1^-$ state
are shown.
The asymptotic values at $R\to \infty$ are subtracted from each energy contribution.
Black dashed lines show 
the two-cluster decay threshold relative to the asymptotic total energy at
$R\to \infty$, given as
$2\epsilon^{(0)}_c-E^{L\pi}_{c+c}(\infty)=-\hbar\omega/4$.
        \label{fig:ene}}
\end{figure*}

\begin{figure}[!htp]
\includegraphics[width=7 cm]{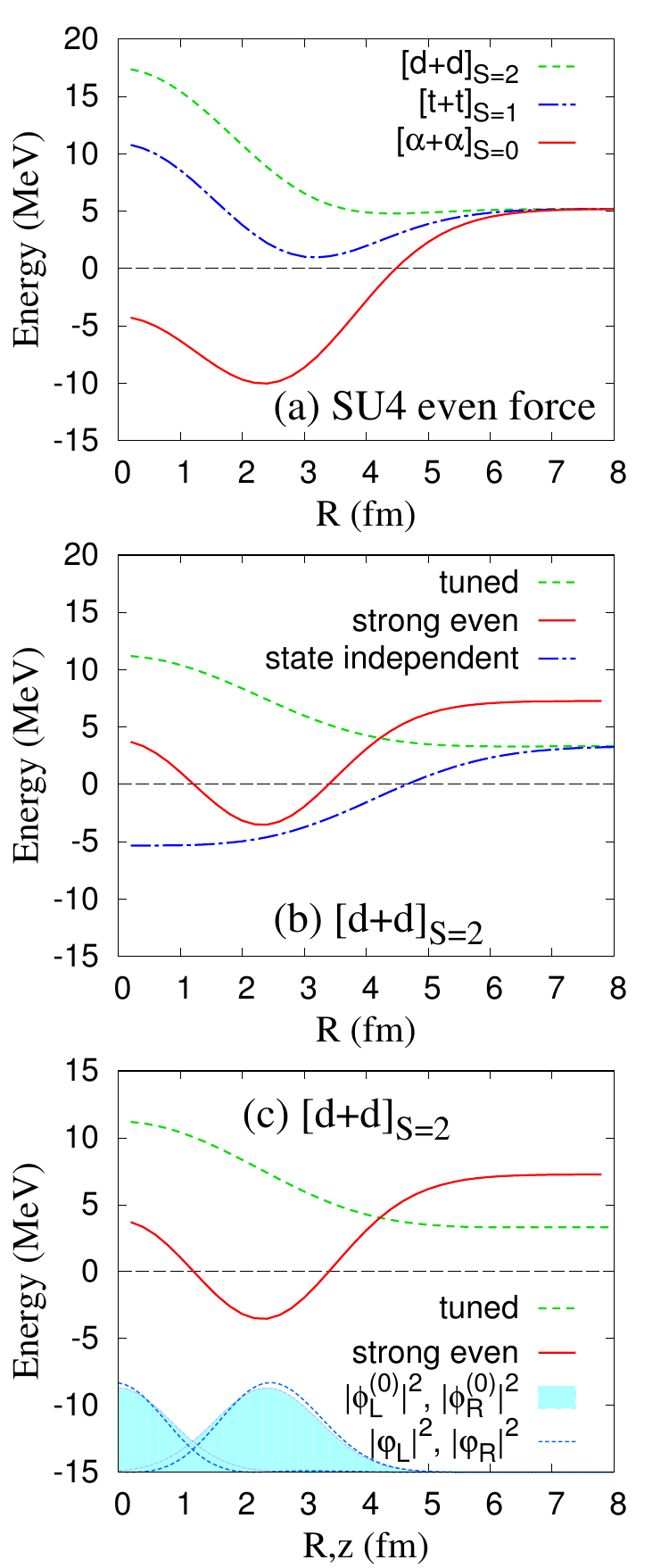}
\caption{
Intrinsic energies $E^\textrm{int}_{c+c}(R)$ 
of (a) $[d+d]_{S=2}$,  $[t+t]_{S=1}$, and $[\alpha+\alpha]_{S=0}$ 
with the SU4-even ($v^\textrm{SU4}_{N}$) force,
(b) $[d+d]_{S=2}$ calculated with the tuned ($v^\textrm{tuned}_{N}$), state-independent
($v^\textrm{st-ind}_{N}$), 
and strong-even ($3v^\textrm{SU4}_{N}$) forces, and those of 
(c) $[d+d]_{S=2}$ calculated with the tuned and strong-even forces. 
The energies measured from the one-dimensional decay threshold energy 
($2\epsilon_c+\frac{1}{2}\hbar\omega$) are plotted. 
In the panel (c), 
the densities of the orthonormal atomic orbitals $\varphi_\newlf(z)$ and $\varphi_\newrt(z)$ 
in the $[d+d]_{S=2}$ system with $\nu=0.35$~fm$^{-2}$ at the distance $R=2.4$~fm
are shown by blue dotted lines in an arbitrary unit, and 
the original atomic orbitals 
$\phi^{(0)}_\newlf(z)$ and $\phi^{(0)}_\newrt(z)$ are shown by 
cyan colored areas. 
        \label{fig:ene-z}}
\end{figure}

To discuss effective inter-cluster interactions, we analyze the
$R$ dependence of the $L^\pi$-projected energies $E^{L\pi}_{c+c}(R)$ for the
two-cluster wave functions $\Phi^{L\pi}_{c+c}(R)$ with the distance $R$.
In Fig.~\ref{fig:ene}, we show the total, kinetic, and 
potential energy contributions of $d+d$, $t+t$, and $\alpha+\alpha$ systems
calculated with the SU4-even ($v^\textrm{SU4}_{N}$) force.
Each energy contribution is shifted by subtracting the
``asymptotic'' value at $R\to \infty$. 
In this plot, the two-cluster decay threshold energy $2\epsilon_{c}$ is located at $\hbar\omega/4$ 
below the ``asymptotic'' total energy at $R=\infty$, which 
contains the kinetic energy cost for localization of the inter-cluster wave function in the $R$ direction.
In Fig.~\ref{fig:ene-z}(a), 
we show the intrinsic energy $E^\textrm{int}_{c+c}(R)$ for $\Phi^{\pi}_{c+c}(R)$ 
measured from the one-dimensional decay threshold energy ($2\epsilon_c+\frac{1}{2}\hbar\omega$). 
Since $\Phi^{\pi}_{c+c}(R)$ 
contains a kinetic energy cost $3(\hbar\omega/4)$ for the localization in three directions, 
the offset energy at $R\to \infty$ is $\hbar\omega/4$.

From the energy curves for $[d+d]_{S=2}$, $[t+t]_{S=1}$, and $[\alpha+\alpha]_{S=0}$, 
the effective inter-cluster interaction in the $[d+d]_{S=2}$ system is found to be repulsive 
for all $R$, whereas those in the $[t+t]_{S=1}$ and $[\alpha+\alpha]_{S=0}$ systems 
are attractive in the medium distance region. 
The kinetic energy term gives a repulsive contribution in the short-distance range because of the Pauli effect, whereas the potential energy term gives an attractive contribution in a slightly longer
range than 
the kinetic repulsion. As the mass number $c$ of clusters increases, the potential energy attraction
rapidly increases, and finally produces the medium-range attraction of the effective interaction  in $[\alpha+\alpha]_{S=0}$. 

In the $[d+d]_{S=0}$ and $[t+t]_{S=0}$ systems, 
the effective inter-cluster interactions are attractive because 
two clusters feel either no or a weaker Pauli effect. In particular, two 
$d$-clusters in the $S=0$ channel feel a rather strong attraction at 
short distances and come close to each other without the Pauli repulsion. 
In these spin-opposed states, two clusters merge into 
bound $\alpha$ and $^6$He states
losing their identity.

In both the spin-aligned and spin-opposed cases, the 
competition between kinetic an potential energy terms plays an important role in the 
effective inter-cluster interactions.
The relatively short-range repulsion of the inter-cluster interactions 
comes from the Pauli effect between identical nucleons
mainly through the kinetic energy term. 

\subsubsection{Energy contributions in the atomic- and molecular-orbital pictures}

For further discussion of the effective inter-cluster interaction 
in the $[d+d]_{S=2}$, $[t+t]_{S=1}$, and $[\alpha+\alpha]_{S=0}$ systems, 
we count kinetic and potential energy contributions
with the atomic- and molecular-orbitals, which are described in the previous section.
For a general discussion, we here choose 
$f_{3E}=f_{1E}\equiv f_\textrm{even}$ and $f_{3O}=f_{1O}\equiv f_\textrm{odd}$
and consider a SU4-symmetric $NN$ force as
\begin{align}
&v_N=V_N(r)(f_\textrm{even}P_\textrm{even}+f_\textrm{odd}P_\textrm{odd}),
\end{align}
with $P_\textrm{even}\equiv P(^3E)+P(^1E)$ and $P_\textrm{odd}\equiv P(^3O)+P(^1O)$.
The energy for a single cluster is given as 
\begin{align}
&\epsilon_{c}=(c-1){\bar T}_{0} + \frac{c(c-1)}{2}{\bar V}_{0}^\textrm{E},\\
&{\bar T}_{0}\equiv\braket{\phi^{(0)}_0|t |\phi^{(0)}_0}=\frac{3}{4}\hbar\omega,\\
&{\bar V}_{0}^\textrm{E}\equiv f_\textrm{even}
\braket{\phi^{(0)}_0\phi^{(0)}_0
  | V_{N}|\phi^{(0)}_0\phi^{(0)}_0},
\end{align}
where $\phi^{(0)}_0=\phi^{(0)}_{\bvec{R}_1=0}$.
Note that the odd component gives no contribution to single-cluster systems. 

For the $[d+d]_{S=2}$, $[t+t]_{S=1}$, and $[\alpha+\alpha]_{S=0}$ systems, the intrinsic energy can be expressed with 
the orthonormal atomic orbitals $\{\varphi_\newlf,\varphi_\newrt\}$ as
\begin{align}\label{eq:ene-atom}
&E^\textrm{int}_{c+c}-2\epsilon_c={\bar T}_{0} +2\Delta\epsilon_c 
+(c^2-c) {\bar V}_{\newlr}^\textrm{E}+c^2{\bar V}_{\newlr}^\textrm{O},\\
& \Delta\epsilon_c= c\bigl({\bar T}_{\newll}-{\bar T}_{0}\bigr)+\frac{c(c-1)}{2}\bigl({\bar V}_{\newll}^\textrm{E}- {\bar V}_{0}^\textrm{E}\bigr),\\
&{\bar T}_{\newll}\equiv\braket{\varphi_\newlf |t |\varphi_\newlf},\nonumber\\
&{\bar V}_{\newll}^\textrm{E}\equiv
 f_\textrm{even} \braket{\varphi_\newlf\varphi_\newlf  | V_{N}|\varphi_\newlf \varphi_\newlf},\nonumber\\
&{\bar V}_{\newlr}^\textrm{E}\equiv  
f_\textrm{even} \braket{\varphi_\newlf\varphi_\newrt  |V_{N}P_\textrm{even}|\varphi_\newlf \varphi_\newrt},\nonumber\\
&{\bar V}_{\newlr}^\textrm{O}\equiv  
f_\textrm{odd} \braket{\varphi_\newlf\varphi_\newrt  |V_{N}P_\textrm{odd}|\varphi_\newlf \varphi_\newrt}.
\end{align}
The first term ${\bar T}_{0}$ 
is the kinetic energy cost to localize two clusters with the distance $R$. 
$\Delta\epsilon_c$ in the second term stands for the internal energy loss of a cluster 
by the cluster distortion from $\{\phi^{(0)}_{\newlf},\phi^{(0)}_{\newrt}\}$ to 
$\{\varphi_{\newlf},\varphi_{\newrt}\}$ because of the Pauli effect. 
The fourth term is the potential energy contribution of the odd part, 
which vanishes for the SU4-even $NN$ force.
The third term for the even part of the potential energy contribution is proportional to 
the factor $c^2-c$ counting the number of different-nucleon pairs. 
In the $[\alpha+\alpha]_{S=0}$ system, this factor is $c^2-c=12$ and 
this third term gives a significant contribution to 
produce the medium-range attraction of the effective inter-cluster interaction,  
whereas the $[d+d]_{S=2}$ system contains only two different-nucleon pairs, which is  
not enough to compensate the repulsion from the first and second terms. 

The energy can be expressed also by molecular orbitals $\{\varphi_+, \varphi_-\}$. 
In the present cluster model, 
$\varphi_+$ and $\varphi_-$ become the harmonic-oscillator $0s$ and $0p$ orbits in the $R\to 0$ limit, 
$\{\varphi_+,\varphi_-\}\to \{ \varphi_s,\varphi_p\}$. In this limit, 
the intrinsic energy of the two-cluster systems are written as 
\begin{align}\label{eq:ene-sp}
&E^\textrm{int}_{c+c}-2\epsilon_c={\bar T}_0+2\Delta \epsilon_c
+(c^2-c) {\bar V}_{sp}^\textrm{E}
+c^2{\bar V}_{sp}^\textrm{O},\\
&\Delta\epsilon_c= \frac{c}{2}\hbar\omega+c({\bar V}_{pp}^\textrm{E}- {\bar V}_{0}^\textrm{E}),\\
&{\bar V}_{pp}^\textrm{E}\equiv f_\textrm{even}\braket{\varphi_p\varphi_p  | V_{N}|\varphi_p \varphi_p},\nonumber\\
&{\bar V}_{sp}^\textrm{E}\equiv f_\textrm{even}\braket{\varphi_s\varphi_p  | V_{N}P_\textrm{even}|\varphi_s\varphi_p},\nonumber\\
&{\bar V}_{sp}^\textrm{O}\equiv f_\textrm{odd}\braket{\varphi_s\varphi_p  | V_{N}P_\textrm{odd} |\varphi_s\varphi_p}.
\end{align}
Here $\varphi_s=\phi^{(0)}_0$ and $\braket{\varphi_p| t|\varphi_p}=5\hbar\omega/4$
are used. 
The internal energy change $\Delta\epsilon_c$ contains the significant repulsive effect from 
the kinetic energy cost for raising half of $A=2c$ nucleons from the $0s$ orbit to the $0p$ orbit 
to avoid Pauli blocking.  

\subsubsection{Bound states of $[d+d]_{S=2}$ with unrealistic $NN$ forces}

As shown in Eq.~\eqref{eq:ene-atom}, 
two $d$-clusters in the $S=2$ channel can be bound 
if the third term $2{\bar V}_{\newlr}^\textrm{E}$  and/or the fourth term 
$4{\bar V}_{\newlr}^\textrm{O}$ 
could give attractive contributions strong enough to compensate 
the kinetic energy increase ${\bar T}_{0}$ and the reduced binding energy of the clusters, 
$2\Delta\epsilon_c$.
We consider two choices corresponding to artificial $NN$ forces which produce a bound $[d+d]_{S=2}$ state. 
One is the strong-even force $3v^\textrm{SU4}_{N}$,
and the other is the state-independent force $v^\textrm{st-ind}_{N}$.
Although these forces do not describe physical nuclear systems, 
it is worth considering these examples in order to better understand the underlying physics involved.

The energies of $[d+d]_{S=2}$ obtained with the GCM and 1d-GCM calculations 
for the $3v^\textrm{SU4}_{N}$ and $v^\textrm{st-ind}_{N}$ forces are shown 
in Table~\ref{tab:two-cluster} together with the deuteron energy $\epsilon_d$, 
and the $R$-plot of the intrinsic energies is shown in Fig.~\ref{fig:ene-z}(b). 

The strong-even force ($3v^\textrm{SU4}_{N}$) gives the deeply bound deuteron cluster
with the cluster size smaller than the range of the $NN$ force, as shown in Fig.~\ref{fig:pot}.
This is in contrast to the tuned $NN$ force $v^\textrm{tune}_{N}$, which provides 
a loosely bound deuteron state with a larger size. 
Moreover, for the deeply bound ``deuteron'' state,  
the potential energy contribution becomes twice the kinetic energy contribution 
(see Table \ref{tab:two-cluster}). 
As seen in the energy curve of the $[d+d]_{S=2}$ system in Fig.~\ref{fig:ene-z}(b), 
a medium-range attraction of the effective inter-cluster interaction is obtained with 
the strong-even force. 
In Fig.~\ref{fig:ene-z}(c), 
we show single-particle densities of the orthonormal atomic orbitals 
in the $[d+d]_{S=2}$ system  
with the distance $R=2.4$~fm to see the cluster distortion due to Pauli effects at intermediate distances.
One can see that left and right atomic orbitals have only small overlap and  
the cluster distortion is minor in this region.
It means that the Pauli repulsion is not crucial in this region, whereas 
the potential attraction $2{\bar V}_{\newlr}^\textrm{E}$ gives significant contribution 
to the binding of the two $d$-clusters. 
In other words, the deeply bound $d$-clusters 
effectively feel a longer-range $NN$ force than the weakly bound $d$-clusters. 
This binding mechanism of the $[d+d]_{S=2}$ system 
is similar to that of a two-dimer system ($Mm+Mm$) with a long-range $Mm$ potential 
previously discussed with the heavy-light ansatz. 

The second case is the state-independent force $v^\textrm{st-ind}_{N}$, 
which contains even-parity and odd-parity components with the same strength. 
It should be commented that 
this $NN$ force is an exactly local potential, whereas other $NN$ forces are 
not but state-dependent forces having no odd-parity component or a weakly repulsive odd-parity component.
The odd component in $v^\textrm{st-ind}_{N}$ force 
gives no contribution to the internal energy of clusters but provides 
an additional attraction to the 
inter-cluster potential.
In Fig.~\ref{fig:ene-z}(b), 
the intrinsic energy of the $[d+d]_{S=2}$ system obtained with the $v^\textrm{st-ind}_{N}$ force is 
shown by a dash-dotted line. The energy curve 
shows an attractive cluster-cluster interaction over a wide range, 
$R\lesssim 5$~fm. 
Different from the case of the strong-even force, there is no short-distance repulsion 
for this case, and the system may go to the $R\to 0$ limit 
with the 
$(0s)^2(0p)^2$ configuration. 

\section{Summary}\label{sec:summary}

We began with a discussion of effective dimer-dimer interactions for general
two-component fermion systems
using the heavy-light ansatz.  In our analysis we were able to give a conceptual understanding of why increasing
the range or strength of the local part of the attractive particle-particle
interaction
results in a more attractive dimer-dimer interaction.     

We then considered the effective cluster-cluster interactions of the $d+d$, $t+t$, and $\alpha+\alpha$ systems using a microscopic cluster model with Brink-Bloch two-cluster wave functions.
As the effective $NN$ force, we use the Volkov central force
with two sets of the parametrization, the SU4-even and tuned $NN$ forces. The latter is 
adjusted to fit the data of the $S$-wave $NN$ scattering lengths 
and the $\alpha$-$\alpha$ scattering phase shifts.
It was shown that the effective inter-cluster interaction in the $[d+d]_{S=2}$ system is repulsive 
for all $R$, whereas those in the $[t+t]_{S=1}$ and $[\alpha+\alpha]_{S=0}$ systems 
are attractive at intermediate distances. 

In these systems,
the kinetic energy term gives a repulsion to the effective inter-cluster interaction
because of Pauli blocking of identical-nucleon pairs.  Meanwhile,
the potential energy term gives an attractive contribution with a slightly longer range than 
the kinetic energy repulsion. As the mass number increases, the potential energy contribution 
increases rapidly and produces enough medium-range attraction to form a bound $2\alpha$ state in the absence of Coulomb effects.
For the $[d+d]_{S=0}$ and $[t+t]_{S=0}$ systems, 
the effective inter-cluster interactions are attractive since the
two clusters feel a weaker Pauli repulsion or none at all.
They then merge to form an $\alpha$ or $^6$He respectively,
giving up their initial two-cluster structures. 

Since the $[d+d]_{S=2}$ system is a two-dimer system of two-component fermions in the isospin sector, the
effective inter-cluster interaction in this system can help to connect with our analysis of
the dimer-dimer interactions for general fermionic systems. 
We extended our analysis of the effective inter-cluster interaction of the $[d+d]_{S=2}$ 
system by artificially changing the $NN$ forces.
It was found that two $d$-clusters could
be bound if two nucleons are deeply bound to form a compact $d$-cluster with a strong even-parity $NN$ force,
or if the $NN$ force contains both even-parity and odd-parity attraction. 

\begin{acknowledgments}  
The computational calculations of this work were performed using the
supercomputer at the Yukawa Institute for Theoretical Physics at Kyoto University. The work was supported
by Grants-in-Aid of the Japan Society for the Promotion of Science (Grant Nos. JP18K03617 and 18H05407), the U.S. Department
of Energy (DE-SC0018638), and the Nuclear Computational
Low-Energy Initiative (NUCLEI) SciDAC project. 
\end{acknowledgments}

\appendix

\section{Two-dimer system with delta potentials in 1D}\label{app:two-delta}
As explained in Sec.~\ref{sec:heavy-light}, the solution for 
the two-dimer system $Mm+Mm$ with a delta $Mm$ potential 
in the heavy-light ansatz ($M\gg m$) is obtained by solving the single-particle problem in the
two-delta potential
\begin{align}
&U(x)=v(x+\frac{R}{2})+v(x-\frac{R}{2}),\\
&v(x)=-\frac{\hbar^2\kappa_0}{m}\delta(x). 
\end{align}
In the frozen dimer ansatz, single-particle energies and wave functions
are approximately expressed as
\begin{align}
&\epsilon_\pm=\epsilon^{(0)}+2\epsilon^{(0)}
\frac{\textrm{e}^{-2\kappa_0 R}\pm \textrm{e}^{-\kappa_0 R}}{1\pm \textrm{e}^{-\kappa_0 R}(1+\kappa_0 R)},\\
&\varphi_\pm(x)=\frac{1}{\sqrt{2[1\pm \textrm{e}^{-\kappa_0 R}(1+\kappa_0 R)]}} \nonumber\\
&\qquad \times\Bigl[\tilde\phi(\kappa_0;x+\frac{R}{2})\pm \tilde\phi(\kappa_0;x-\frac{R}{2})\Bigr],
\label{eq:app-phi}\\
&\tilde\phi(\kappa;x)=\sqrt{\kappa}\textrm{e}^{-\kappa|x|}, 
\end{align}
where $\epsilon^{(0)}=-\frac{\hbar^2}{2m}\kappa^2_0$ and 
$\tilde\phi(\kappa_0;x)=\phi^{(0)}(x)$ are the single-particle energy and wave function for 
the bound-state solution in
the single-delta potential $U(x)=v(x)$. 
$\varphi_+(x)$ and $\varphi_-(x)$ are the molecular orbitals with positive and 
negative parities. 

For the exact energies $\epsilon^\textrm{exact}_\pm$, we define valuables 
\begin{align}
\kappa_\pm=\frac{\sqrt{-2m \epsilon^\textrm{exact}_\pm}}{{\hbar}}.
\end{align}
$\kappa_\pm$ satisfy equations
\begin{align}\label{eq:kappa}
\frac{\kappa_\pm}{\kappa_0}=1\pm \textrm{e}^{-\kappa_\pm R},
\end{align}
and the solutions are given 
as 
\begin{align}
&\kappa_+=\kappa_0\Bigl\{1+\frac{1}{\kappa_0R}W_0(\kappa_0R \textrm{e}^{-\kappa_0 R})\Bigr\}, \\
&\kappa_-=\kappa_0\Bigl\{1+\frac{1}{\kappa_0R}W_{-1}(-\kappa_0R \textrm{e}^{-\kappa_0 R})\Bigr\}.
\end{align}
The exact single-particle energies and wave functions 
are written with $\kappa_\pm$ as 
\begin{align}
&\epsilon^\textrm{exact}_\pm= -\frac{\hbar^2}{2m}\kappa^2_\pm,\\
&\psi_\pm(x)=\frac{1}{\sqrt{2[1\pm \textrm{e}^{-\kappa_\pm R}(1+\kappa_\pm R)]}} \nonumber\\
&\qquad \times\Bigl[\phi(\kappa_\pm;x+\frac{R}{2})\pm \phi(\kappa_\pm;x-\frac{R}{2})\Bigr].
\label{eq:exact-phi}
\end{align}
Note that the negative-parity state is not bound for $\kappa_0 R<1$, meaning that
the two-delta potential is not enough 
to bind two fermions.

By comparing Eqs.~\eqref{eq:app-phi} and \eqref{eq:exact-phi}, 
one can see that the approximate single-particle wave functions $\varphi_\pm(x)$
are expressed in a similar form to $\psi_\pm(x)$, but $\kappa_0$ for the 
unperturbed energy $\epsilon^{(0)}$ is used in $\varphi_\pm(x)$ 
instead of $\kappa_\pm$ for the exact solutions. 

Also for the two-delta potential in 3D, exact single-particle energies 
($\epsilon^\textrm{exact}_\pm= -\frac{\hbar^2}{2m}\kappa^2_\pm$)  
for the positive-and negative-parity 
bound states can be expressed in similar forms 
with $\kappa_\pm$  
given in Eq.~\eqref{eq:kappaW-3D}, which satisfy equations
\begin{align}\label{eq:kappa-3D}
\frac{\kappa_\pm}{\kappa_0}=1\pm \frac{\textrm{e}^{-\kappa_\pm R}}{\kappa_0 R}.
\end{align}
The full details for the 3D case can obtained from the authors upon request.

\section{Effective $NN$ interaction} \label{app:volkov}

The effective $NN$ force $v_{N}(i,j)$
used in the present calculations of two-cluster systems is the Volkov central force \cite{VOLKOV}, which is a finite-range two-body nuclear force with a Gaussian form as
\begin{align}
&v_{N}(1,2)\nonumber\\
&\qquad =V_{N}(r)(W+BP^\sigma_{12}-HP^\tau_{12}-MP^\sigma_{12} P^\tau_{12}),\\
&V_{N}(r)=\sum_{k=1,2} V_k e^{-\frac{r^2}{\eta_k^2}},\qquad 
r\equiv \sqrt{\bvec{r}_2-\bvec{r}_1},
\end{align}
where $P^\sigma_{12}$ and $P^\tau_{12}$ are the exchange operators of 
nucleon-spins and isospins, respectively. For the strength and range parameters, 
we use the Volkov No.2 parametrization given as $V_1=-60.65$~MeV,
$V_2=61.14$~MeV, $\eta_1=1.80$~fm, and $\eta_2=1.01$~fm. 

The Volkov $NN$ force can be rewritten as, 
\begin{align}
v_{N}(1,2)&=V_{N}(r)\bigl[f_{3E}P(^3E)+ f_{1E}P(^1E)\nonumber\\
&+f_{3O}P(^3O)+f_{1O}P(^1O)\bigr],
\end{align}
with
\begin{align}
f_{3E}=W+B+H+M, \nonumber\\
f_{1E}=W-B-H+M, \nonumber\\
f_{3O}=W-B+H-M, \nonumber\\
f_{1O}=W+B-H-M.
\end{align}
It means that the strengths of the 
$^1E$, $^3E$, $^1O$, and  $^1O$ terms can be adjusted by 
$W$ $B$, $H$, and $M$ for the Wigner, Bartlett, Heisenberg, and Majorana 
terms, respectively, in the Volkov force. 

The parameter sets of $W$, $B$, $H$, and $M$ for 
the SU4-even, tuned, strong-even, and state-independent forces
used in the present calculation and corresponding values of 
$f_{3E}$, $f_{1E}$, $f_{3O}$, and $f_{1O}$ are 
summarized in Table \ref{tab:volkov}.

The $v^\textrm{tuned}_{N}$ force is adjusted to fit the experimental data of the  
$S$-wave $NN$ scattering lengths $a_t$ in the spin-triplet and $a_s$ in the spin-singlet,
and also the $\alpha$-$\alpha$ scattering phase shifts.
The theoretical values obtained with the $v^\textrm{tuned}_{N}$ force are 
$a_t=5.4$~fm and $a_s=-23.9$~fm, and the experimental values measured by 
$pn$ scattering are $a_t=5.42$~fm and $a_s=-23.75$~fm ~\cite{Dumbrajs:1983jd}.

\section{Relative wave function between clusters in two-cluster wave functions} \label{app:relative-wf}
The spacial part of the two-cluster wave function $\Phi_{c+c}(\bvec{R})$ in Eq.~\eqref{eq:ccatomic}
can be rewritten in a separable form of the cm, inter-cluster, and intrinsic coordinates as 
\begin{align}
&\Phi^{c}_{-\frac{\bvec{R}}{2}}({\bvec{r}}_1,\ldots,{\bvec{r}}_c) 
\Phi^{c}_{\frac{\bvec{R}}{2}}({\bvec{r}}_{1'},\ldots,{\bvec{r}}_{c'})
\nonumber\\
&=\phi_\textrm{cm}(\bvec{r}_\textrm{cm})\otimes\phi_\textrm{rel}(\bvec{R},\bvec{r}_\textrm{rel})
\otimes\Phi^c(\bvec{\xi})\otimes\Phi^c(\bvec{\xi}')\nonumber\\
&\quad  \otimes\bigl[\chi_c(s_1,\ldots,s_c)\chi_c(s_{1'},\ldots,s_{c'})\bigr]_{S},
\label{eq:cm-rela}\\
&\phi_\textrm{cm}=\left( \frac{4c\nu}{\pi} \right) e^{-2c\nu\bvec{r}_\textrm{cm}^2},\\
&\phi_\textrm{rel}(\bvec{R}, \bvec{r}_\textrm{rel})= \left ( \frac{2\gamma}{\pi}\right )^{3/4}
e^{-\gamma(\bvec{r}_\textrm{rel}-\bvec{R})^2} \nonumber\\
&=\sum_L\Gamma_L(r_\textrm{rel},R)\sum_m 
Y_{LM}(\hat{\bvec{r}}_\textrm{rel})Y^*_{LM}(\hat{\bvec{R}}),\\
& \Gamma_L(r_\textrm{rel},R) \equiv 4\pi (\frac{2\gamma}{\pi})^{\frac{3}{4}} i_L(2\gamma R r_\textrm{rel}) e^{-\gamma(r_\textrm{rel}^2+R^2)},\label{eq:Gamma}\\
&\bvec{r}_\textrm{cm}\equiv \frac{1}{2c} \sum_{i=1}^c(\bvec{r}_i+\bvec{r}_{i'}), \quad \bvec{r}_\textrm{rel}\equiv 
\frac{1}{c}\sum_{i=1}^c (\bvec{r}_i-\bvec{r}_{i'}),
\end{align}
with $\gamma \equiv  \frac{c}{2} \nu$. Here
$\bvec{r}_\textrm{cm}$, $\bvec{r}_\textrm{rel}$, $\bvec{\xi}$, and $\bvec{\xi}'$
indicate the cm coordinate, 
the inter-cluster coordinate, and intrinsic coordinates of the first and second clusters,
respectively.

The GCM calculation is performed
by superposing $L^\pi$-projected wave functions as given in Eq.~\eqref{eq:gcm-BB}.
The GCM calculation is equivalent to
optimization of the inter-cluster wave function by the expansion 
with the base function 
$\Gamma_L(r_\textrm{rel},R_k)$ as
\begin{align}
&\Psi^{\textrm{GCM}}_{c+c}\nonumber \\
&={\cal A}\Bigl\{
\phi_\textrm{cm}(\bvec{r}_\textrm{cm})\otimes\psi^\textrm{GCM}(r_\textrm{rel})Y_{LM}(\hat{\bvec{r}}_\textrm{rel})
\otimes\Phi^c(\bvec{\xi})\otimes\Phi^c(\bvec{\xi}') \nonumber\\
&\quad \otimes\bigl[\chi_c(s_1,\ldots,s_c)\chi_c(s_{1'},\ldots,s_{c'})\bigr]_{S} 
\Bigr\},\\
&\psi^\textrm{GCM}(r_\textrm{rel})=\sum_k c_k \sqrt{\frac{2L+1}{4\pi}}\Gamma_L(r_\textrm{rel},R_k). 
\end{align}
Similarly, the inter-cluster 
wave function $\psi^\textrm{1d-GCM}(\bvec{r}_\textrm{rel})$ in the 1d-GCM wave function 
is given as 
\begin{align}
&\Psi^{\textrm{1d-GCM}}_{c+c}\nonumber\\
&={\cal A}\Bigl\{
\phi_\textrm{cm}(\bvec{r}_\textrm{cm})\otimes\psi^\textrm{1d-GCM}(\bvec{r}_\textrm{rel})
\otimes\Phi^c(\bvec{\xi})\otimes\Phi^c(\bvec{\xi}') \nonumber\\
&\quad \otimes\bigl[\chi_c(s_1,\ldots,s_c)\chi_c(s_{1'},\ldots,s_{c'})\bigr]_{S} 
\Bigr\},\\
&\psi^\textrm{1d-GCM}(\bvec{r}_\textrm{rel})=\sum_k c_k P^\pi\phi_\textrm{rel}(\bvec{r}_\textrm{rel},\bvec{R}_k), 
\end{align}
with $\bvec{R}_k=(0,0,R_k)$.

\end{document}